\documentclass[12pt]{article}
\usepackage{graphicx}

\begin{document}
\begin{center}
\begin{huge}
Cont-Bouchaud percolation model including Tobin tax\\[1.0cm]
\end{huge}
Gudrun Ehrenstein\footnote{email:ge@thp.uni-koeln.de}\\
Mai 15, 2002\\
Institute for Theoretical Physics,Cologne University,50923 K\"oln\\
 Euroland\\[1.0cm]
\end{center}
\textsl{Abstract:\\
The Tobin tax is an often discussed method to tame speculation and get
a source of income. 
The discussion is especially heated when the financial markets are 
in crisis. In this article we refer to foreign exchange markets.
The Tobin tax should be a small international tax affecting
 all currency transactions and thus consequently reducing the destabilizing 
speculations. In this way this tax should take  over a control 
function. By including Tobin tax in the microscopic model of 
Cont and Bouchaud one finds that Tobin tax could be the right 
method to control foreign exchange operations and get a good source of income.}\\[1.0cm]

\section{Introduction}
The reader who is familiar with the economically background of
 foreign exchange markets may skip this section.

\subsection{Financial markets in particular the foreign exchange 
markets}
The financial markets are composed of credit markets, security 
markets and foreign exchange markets. These three markets interact. 
In this article we refer to foreign exchange markets.
On foreign exchange markets currencies are traded. Exchange rate 
 changes are a measure for the economical efficiency of national 
economics. The rate of exchange is the price for a foreign currency.
 Increasing or decreasing prices of the currencies affect the 
development of the particular national economy. So a devaluation 
causes expensive imports and low priced exports.\\

\subsection{The system of Bretton Woods}
The foreign exchange market we know today existed not before the seventies
 in this way. Fourty years ago, the system of Bretton Woods reigned 
over the international financial world. In 1944 agents of 44 nations 
met in Bretton Woods in the US-state New Hampshire under leadership 
of the victor nations to decide about the economic future after the 
World War II. The members of the conference brought about the World 
Bank and the International Monetary Fund. The arrangement of Bretton 
Woods was enacted in December 1945. The purpose of this meeting was to 
reorganize and to stabilize world commerce and international trade 
after the Second World War. The new world currency system is based on 
the warranty of best possible free convertibility of the currencies 
with fixed foreign exchange rates. The member states had to trade 
their currencies either for parities in gold or parities to the
US Dollar, which showed a gold parity itself. 
So the US Dollar became the new world's leading currency [1].\\

\subsection{Foreign exchange markets after Bretton Woods}
The system of Bretton Woods was exempted at first from international
 financial crises. In Germany we know the time of economic growth as 
``economic miracle''. The arrangement of Bretton Woods was only 
profitable for industrial nations. For developing countries  
the promised wealth was never reached. Later, the weakening strength of
the US Dollar compared to European currencies lead to speculative crises.\\

Thus since 1971 the arrangement was abolished step by step.
 For reduction of the great nonequilibrium in the balances of 
payments between the EG currencies and the US currency, free 
rates of exchange were introduced in 1973. The fixed foreign 
exchange rates were replaced by free fluctuating foreign exchange 
rates between the leading currencies which are today US Dollar, 
Euro and Yen. The rate of exchange system was regulated by the 
free game of the market forces, e.g.the private finance companies.
 The government did not interfere in foreign exchange market regularly.
 The prices for different currencies result from the difference 
between demand and supply. Important 
for the rate of exchange is the faith of the investor in the currency.
 Since controls for the turnover of capitals no longer exist and
 computers render it possible to do transactions very quickly,
 rates of exchanges are determined by short-term financing. 
Consequently, the foreign exchange markets are not a true measure 
for the economical efficiency of the nations [2].

\subsection{Where does the trade with foreign exchanges take place ?}
62\% of the world wide trading volume with foreign exchange in 1995 
was realized from only a few finance companies in the top five markets
: United Kingdom, United States, Japan, Singapore and Hong Kong. 
85\% of this trade takes place in the top nine (top five plus 
Switzerland, Germany, France and Australia) alone [2,4].\\

\subsection{Crises}
Unfortunately the less established financial markets today are 
determined by proneness  to crises (financial crises in the nineties and later:
 1994 Mexico, 1997 East Asia, 1998 Russia, 1999 Brazil [4] and 2001/2002
Argentina.). Short-term transactions are in a high
 measure responsible for the fluctuations of the rates of exchanges.
 Volatility favours the development of speculative bubbles and in this
 way it could give rise to crises. This is especially difficult for
 developing countries because they react very sensitively on external
 crises.\\
A model of financial crises is a four step one by  Kindleberger[2]:\\

First stage:\\
An external occurence promises opportunities for making profit. 
The price for the hopefully profitable thing increases.\\

Second stage:\\
This attracts other investors, who want to profit from the increasing 
prices. The increasing demand causes an increase of the price. The 
higher the expectation of profit the more money will be invested. 
If the investors believe in very high profit they raise a loan to invest 
more in the speculative operation.\\

Third stage:\\
The investors come to know that the profit expectation can not be fulfilled. 
So the profit expectations turn back. The prices increase only slowly 
or they remain stable.\\

Last stage:\\
The prices stagnate and the investors who financed the speculative 
operation by credits can't pay back the credits. They have to give up 
all or parts of the securities. The prices of the speculative object 
decrease. The decreasing prices cause investors to sell
 their papers panickingly. The worth of the speculative object goes down.
 So profit expectation turns into losing business and credits can't be 
payed back. Papers which are used as securities for further credits lose
 on worth and the banking establishment calls for further securities, 
repay or they call in the credits. A bank crisis is possible. The credit 
crisis turns to an insolvency of the inland economy. So the firms are no
 longer solvent. The financial crisis is now in production sector, and 
this causes unemployment with its consequences.\\[0.2cm]
 World wide, the ILO, 
in its 1998 World Employment Report, estimated that unemployment 
increased by 10 million people solely due to the Asian financial crises [3]. 
Often such crises spread out internationally. Above all short-term
 capital played an important role in the finance crises of 
the nineties. Their part on all trades increased about by 300 
percent between 1990 and 1995. 
An important reason for this growth is the removal of controls over 
turnover of capitals. By means of such controls long-term credits or
 capital investments could be preferred over short-term investions. 
Additionally, interests for short-term credits are more favourable 
than those for long-term credits, because the risk for creditors 
is greater with long-term credits for developing countries.\\

\subsection{How does currency speculation work ?}
If an investor expects the devaluation of a currency A, 
he will raise a credit in this currency A. The money he gets in this way
 will be invested in a currency B. When the devaluation of currency A occurs,
 the investor only has to take a smaller part of his money 
in currency B to pay back the credit in currency A. The rest of the money 
in currency B is his profit.\\

\subsection{Herd behaviour}
Through hedge funds and other financial organisations which operate 
internationally the speculation on currencies gets more weight. When 
a big financial organisation invests a lot of money or calls a lot of
 money back from a certain transaction, profit or loss expectation from
 other investors rises. Thus a herd behaviour springs up. A lot of 
investors will try to get a part of the quickly earned money. Much money will
 be  invested in speculative operations. Sometimes the foreign exchange 
reserve of the central bank will be spent, if the central bank tries 
to stabilize the own currency by means of buying. If this method fails,
 the currency speculation is followed by a devaluation of the currency.\\

\subsection{Trading volume on foreign exchange markets}
When in 1971 the USA abandoned the Bretton Woods system of fixed foreign 
exchange rates and the first transactions with computers came into being,
 the turnover on the financial markets grew in an abnormal way. 
From 1970 to 2000 the turnover of the trade with currencies increased
 from 70 billion to 1.5 trillion US Dollars each trading day. Especially 
80\% of the 1.5 trillion Dollars are short-term transactions with terms 
less than seven days, and more than 40\% involve round-trips (a purchase 
operation followed by a resale operation) within two days or less [3]. 
To estimate the dimensions of speculation we have to take into 
consideration that annual turnover by 200 trading days in a year 
reaches 300 trillion US Dollar 
and this is 40 times the international product and service trade. 
And 90\% of the turnover are speculative short-term transactions and 
only 10\% goes into the production sector [5]. 

\subsection{The Tobin tax}
To slow down this process in order that less operations will be 
speculative and the rates of exchange less fluctuating, James 
Tobin\footnote{James Tobin was an American economist. Since 
1955 he taught at Yale University. In 1981 Tobin became the Nobel
 Laureate for his papers about monetary theory for governmental 
financial management [5].}
 already suggested in 1972 to impose a 
tax with a tax rate of 1\% affecting all buying and selling currency 
transactions\footnote{Tobin's proposed tax on international currency 
transactions, intended to curb speculation, is an extension to foreign 
exchange markets of Keynes's proposed tax on stock market transactions. 
Keynes advocated a tax to curb speculation in stock markets. He advocated 
stronger measures for foreign exchange markets, such as capital controls 
to defend the autonomy of national stabilization policies. These capital 
controls were an essential factor to Keynes's wartime proposal for an 
International Clearing Union [4].}. 
Today a tax rate between 0.05\% and 
0.5\% is discussed. The tax Tobin recommended should affect the 
speculators. The investors try to use smallest price differences 
even below the $10^{-3}$ border by foreign rates of exchange.
 This transactions don't have a real economic 
meaning. The development of the exchange rates reflect only the hopes 
of the investors. Supporters of the tax believe that a small tax will 
make such transactions unprofitable. The rate and the number of such 
short-term transactions slow down without affecting long-term credits 
and long-term capital investment. More precisely, they talk about a filter 
function of the Tobin tax: The tax renders all currency transactions 
( no matter whether short-term or long-term) more expensive. But because 
of the different terms the consequences for the different dealing are
very different. Short-term transactions are 
only lucrative when the expectation of the profit is higher than Tobin tax. 
For example: if Tobin tax rate is 0.5\% and one wants to do a transaction 
for only one week: in this case an annual interest of 52\% is necessary 
to make the transaction profitable. The reduction of short-term transactions 
will favour long-term transactions. Speculation is reduced and this 
is hoped to be a prevention for crises. But there is still a fact 
where Tobin tax 
is not a method to control. Even the supporters of Tobin tax says that 
this kind of tax can't minimize speculation operations where 10\% to 
50\% profit is promised. Therefore other controls about turnover of 
capital should be used. Supporters of the tax named two reasons for 
introducing Tobin tax: on the one hand the tax would regulate the 
speculative operations, on the other hand there will be a source of 
income which is estimated in a two figure billion US Dollars range. 
Estimates from Felix and Sau in 1996 tell that a tax rate of 0.25\% 
will reduce the transaction volume less than 33\% [5]. With these 
reasons for Tobin Tax in mind, many people in government
 or from other organisations 
like WEED or ATTAC\footnote{On initiative of the french monthly ``
Le Monde diplomatiqu\'e'' come out in june 1998 ATTAC ``Association 
pour une taxation financieres pour l'aide aux citoyens''. ATTAC is 
now represented in 30 countries. A purpose from ATTAC is that policy
regulates financial markets and not reverse.} agree to the tax.\\

\section{The model of Cont and Bouchaud [6,7]}
We get our results due to combining Tobin tax in two different ways 
with the microscopic 
model of Cont and Bouchaud based on percolation theory.\\
 In percolation theory we start to fill 
the lattice in the way that each site is randomly occupied with 
probability $p$ and empty with probability $(1-p)$. Neighboring occupied 
sites form clusters. If a contiguous path of occupied sites 
 connects upper and bottom of the lattice for the first time, 
the threshold value $p=p_c$ is reached.\\
In the model presented, the randomly formed clusters are agents who act 
together, to describe herding behaviour
 in financial markets. Different clusters reach their decision about 
buying or selling  absolutely randomly and indepently from the other 
clusters. If there is more supply than demand, the price change decreases
 proportionally to the difference -  otherwise the price change increases.
In every iteration, each cluster makes a decision to buy or to sell,
 each with probability $a$ (which is called the activity). The cluster 
is inactive in the iteration with probability $(1-2a)$. The maximum value
 of the activity is 0.5 . We can interpret the activity as a measure 
for length of the time which we handle in one iteration. If we take 
the value $a$ close to 0.5, we handle big time steps where nearly all 
clusters have already made a decision to buy or to sell. Otherwise 
a small $a$ means that one iteration represents not enough time so that 
there are only a few clusters who make in this iteration a decision 
to buy or to sell.\\
The probability distribution of the price fluctuations shows 
many smaller and less big fluctuations and the graph is essentially 
symmetrical. For an activity $a$ close to 0.5 the results at the 
critical point $p_c$ are similar to a Gaussian curve. For a smaller 
activity (in particular if we take an $a$ corresponding to intra day 
time scales) we get heavy tails in the distribution of stock price 
variations in the form of a power law truncated by size effects. Such 
a behaviour was observed in empirical studies of high frequency market 
data.  Furthermore there exists weak correlations between 
successive price changes, and 
strong correlations (``volatility clustering'') between successive 
absolute values of price changes.\\
\begin{center}
\textsl{The framework of the simulation}
\end{center}
\begin{itemize}
\item We determine with the complicated algorithm of Hoshen and Kopelman 
the number $n_s$ of clusters with $s$ agents. Note that if one works with 
$p$ greater than $p_c$, one has to ignore the one infinite cluster. This 
infinite cluster causes only crashes and bubbles.
\item We decide randomly if the cluster is active in this iteration.
\begin{itemize}
\item If the cluster is active, we decide by another random number if 
the cluster would like to buy or to sell an amount which corresponds
 to the size of the cluster. The return, the 
difference $\sum_s({n_s^+*s-n_s^-*s})$ which means especially the difference 
between demand and supply, is proportional to price change in this time step.
\item If the cluster is not active, it contributes nothing to the return.
\end{itemize}
\item If we have all clusters processed in this way, one iteration 
is finished and we start again from beginning.
\end{itemize}
The best agreement with real price fluctuations is found when the 
concentration is slightly above $p_c$. One gets the biggest fluctuations 
when $p$ is equal to $p_c$. Cont and Bouchaud identified $p=p_c$ therefore 
with market crashes.\\
There are already different modifications of this model in existence [8].

\section{Modifications of the model of Cont and Bouchaud by 
introducing Tobin tax}
In the original Cont-Bouchaud model we have only two parameters: the 
activity $a$ and the concentration $p$ (the probability to find an occupied 
site in the lattice. In our simulations we work always at the critical 
point $p_c$). The modified versions of the model used here, has three 
additional parameters: Tobin tax, Producers and maxwin.\\
 Producers are people who act, even when they not expect to 
make profit, e.g. to pay a bill.\\
 Because the returns of the Cont-Bouchaud model are often large integers 
and the Tobin tax is a very small number (less than 1\%), we have to 
normalize the returns. We take two values for maxwin here: 50\% and 5\%.
 In the first case, this means that if all clusters in an iteration 
are active and buying, the return is +50\%. 
Otherwise, if all clusters are active and selling only, the return
 is -50\%. Certainly, not all clusters will buy in the 
same iteration and not all clusters will sell in the same iteration. 
 We actually get returns between -8\% and +8\% with the 
original Cont-Bouchaud model in the 
case of maxwin= 50\%. For the case of maxwin= 5\% we get actual returns
 between -0.8\% and +0.8\% (see figures 8a, 8b and 9). Note
 that 80\% of the daily speculation trade has taken place because the 
traders would like to take advantage of profits below the $10^{-3}$ border.
 Thus the value maxwin= 5\% is rational.\\ 

\subsection{First modification of the Cont-Bouchaud model} 
\begin{center}
\textsl{The framework of the simulation}
\end{center}
\begin{itemize}
\item Because our agents believe that the return of the time step 
before, which we call $r_-$, 
is authoritative for the following return development, we need $r_-$ 
in this model. In the first time step we took $r_-$= 1\%.
\item We determine the number $n_s$ of clusters with $s$
 agents with the algorithm of Hoshen and Kopelman.
\item We decide randomly if the cluster is active in this iteration.
\begin{itemize}
\item If the cluster is active we have to test, whether the condition 
($r_-$ greater than Tobin tax)
 has been fulfilled or not.\\
\begin{itemize}
\item If this is the case we decide by another random number, if the 
cluster would like to buy or to sell an amount which corresponds to the 
size of the cluster.\\
Notice: Both, for buying and selling, we take the condition $r_-$ greater 
than Tobin tax. In the first case we simulate optimists (these are people 
who buy because $r_-$ is greater than Tobin tax and they believe that $r_-$ 
of next time step will exceed current one) and in the second pessimists 
(these people sell because $r_-$ is greater than Tobin tax and they believe 
that $r_-$ in the next time step will be in the same order of magnitude as 
in the current time step but it will change the sign).\\
 The return, the difference $\sum_s({n_s^+s-n_s^-s})$ which means the difference 
between demand and supply, is proportional to the price change in this time step. 
\item If the condition is not true we decide through another random number 
if the cluster is forced to trade because it belongs to the Producers.
\begin{itemize}
\item If the cluster is a Producer we decide randomly if the cluster
 buys or sells and take it into account for the calculation of the return.
\item If the cluster does not trade it contributes nothing to return 
or turnover.
\end{itemize}
\end{itemize}
\item  If the cluster is not active it contributes nothing to return 
or turnover.
\end{itemize}
\item At the end of the iteration we save the return 
to use it in the next iteration as new $r_-$. We normalize the returns as 
described above.
\item If we have handled all the clusters in this way, we have finished 
one time step and we begin the next by redistributing anew the occupied sites.
\end{itemize}

Note that when there are no Producers and the condition is never fulfilled 
the return is always zero.\\

\subsection{Second modification of the Cont-Bouchaud model}
\begin{center}
\textsl{The framework of the simulation}
\end{center}
\begin{itemize}
\item We took again $r_-$= 1\% initially.
\item We determine the number $n_s$ of clusters with $s$ agents.
\item We decide randomly if the cluster is active in this iteration.
\begin{itemize}
\item If the cluster is active we have to test, whether one of the conditions 
($r_-$ greater equal than Tobin tax or $r_-$ less than ($-$ Tobin tax))
 has been fulfilled or not.\\
\begin{itemize}
\item If this is the case we decide by another random number, if the cluster 
would like to buy or to sell an amount which corresponds to the size of the cluster. 
The cluster trades in the following two cases: 1. $r_-$ greater than Tobin tax 
and the cluster will buy (this simulates the behaviour of optimists because they 
believe that the return will further increase). 2. $r_-$ less than ($-$ Tobin tax) 
and the cluster will sell (this simulates the behaviour of pessimists because they 
believe that the return will further decrease). As above, we determine the return 
by the difference between demand and supply.\\
\end{itemize}
\end{itemize}
\item Now for Producers etc. we procede as in the first modification. 
\end{itemize}
We examine the second modification only for RM.

\section{Results}

 To get figures 1-6,11-13 and 16, we took for our calculations an average over 
about 10000 cluster 
configurations and a square lattice of length 31.
 The activity was a=0.4999.\\
For many of our results we calculate the turnover and sum it up over 
500 time steps. With this turnover we determine the profit for government 
as a function of Tobin tax by multiplication of the turnover with the Tobin tax rate. 
To get better results we multiply profit for government by $10^{8}$, which corresponds 
to a summation of turnover about more iterations. The model without Tobin tax can be 
related to Tobin tax equal zero (which we call ``rational model'' (RM)) or 
alternatively to the original Cont-Bouchaud model, where all investors trade even for 
negative $r_-$ (which we call ``similar Cont-Bouchaud model (SCBM)). When we examine 
y as a function of Tobin tax, the differentiation between these two models leads to an 
other scale of the y axis. To see this, we  show the comparison between these two 
models in figures 3 - 6 for the first modification of the Cont-Bouchaud model. We 
examine the second modification only for RM.\\

 To get the turnover as a function of Tobin tax, we set the turnover we get the way 
described above for zero Tobin tax to 100\%. Then we relate the turnover we get for a 
positive Tobin tax to the one when we set the Tobin tax to zero.\\
(Alternatively, we could normalize here again the turnover to that of the original 
Cont-Bouchaud model.)

\subsection{Results of the first modification of the Cont-Bouchaud model}
\begin{itemize}
\item Turnover as a function of time\\[0.2cm]
In fig.1a we see that trade will go to zero when there are no Producers.
We also see the way the trade volume will decrease with increasing 
Tobin tax when there are 0.5\% Producers. We took maxwin= 5\%\\[0.1cm]
The graph shows from top to bottom:\\
The first line shows the turnover we get with the original Cont-Bouchaud 
model. The next line is a turnover when  people only trade, if $r_-$ is greater 
or equal zero.\\
 We get the following three by introducing 0.5\% Producers with a Tobin tax of zero 
(empty circles), 0.16\% (full circles) and 0.5\% (triangles).
 The next three 
lines show the development of turnover when nobody has to act, 
i.e. when there are no Producers, for the same three 
tax levels.\\[0.1cm]
What we see if there are no Producers, the turnover approaches 
zero the faster, the higher Tobin tax is.\\[0.1cm]
\begin{figure}[!]
\begin{center}
\includegraphics[angle=-90,scale=0.31]{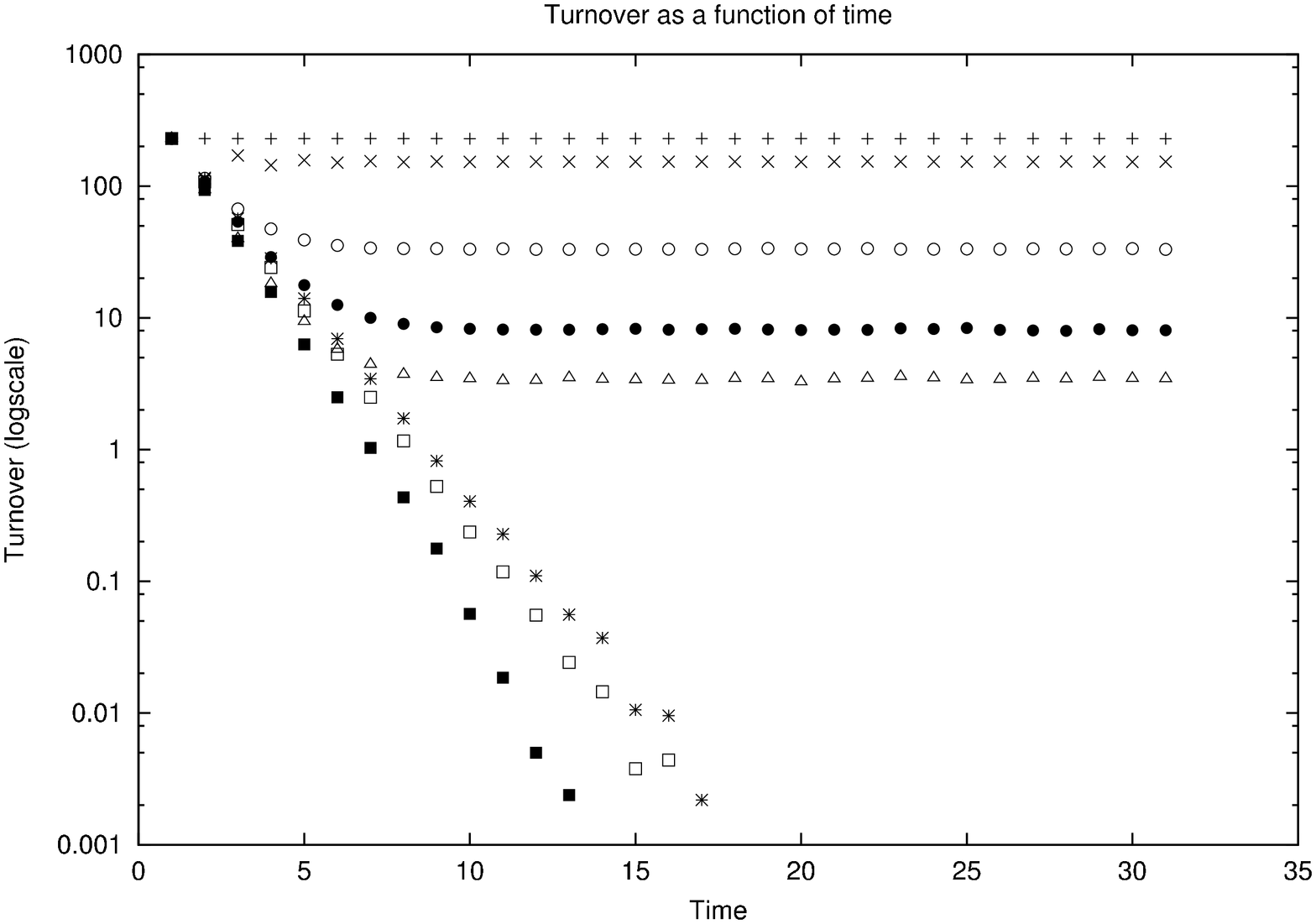}
\includegraphics[angle=-90,scale=0.31]{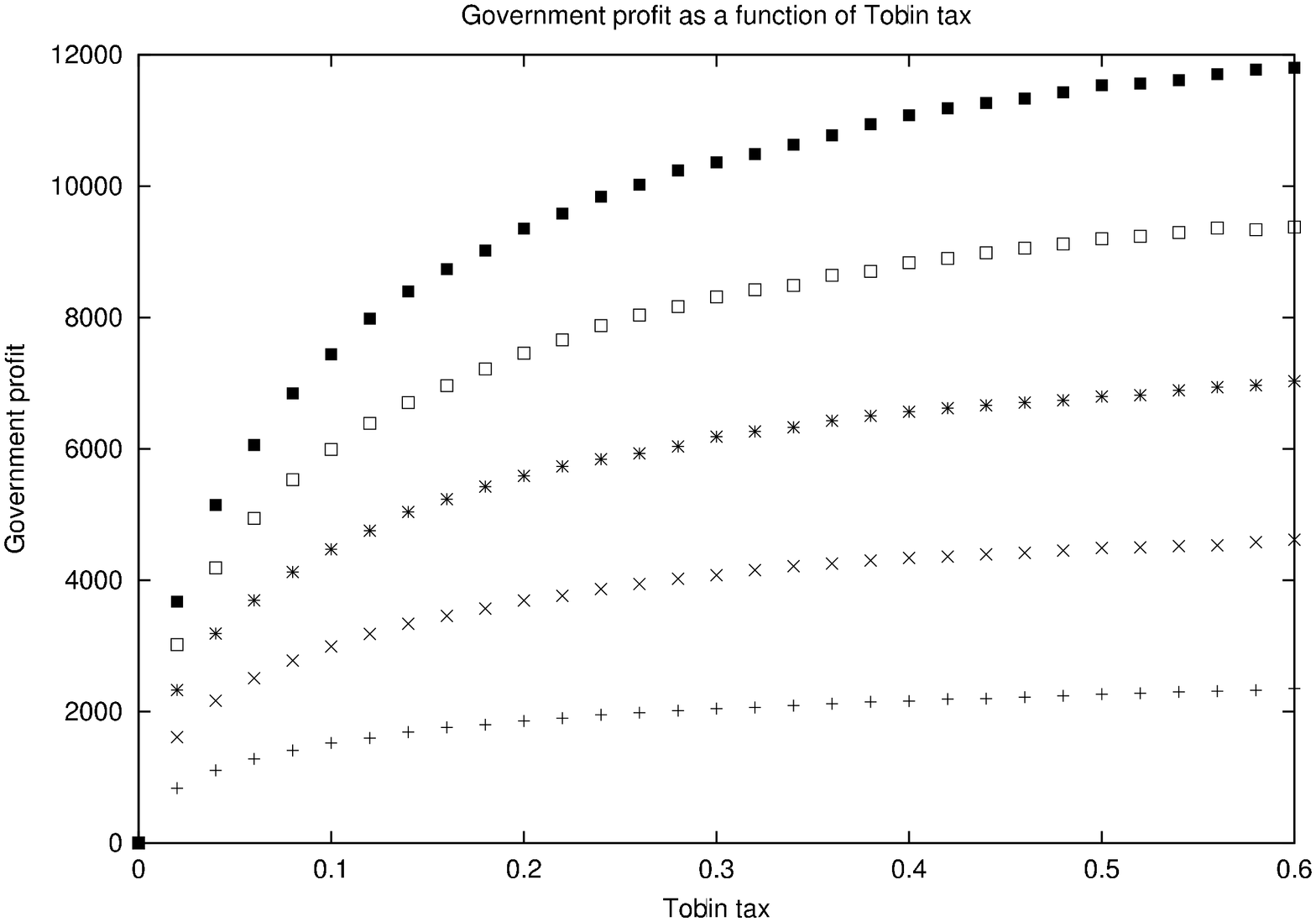}
\end{center}
\caption{
a) Turnover as a function of time. b) Government profit as a function of  
Tobin tax. From top to bottom: 0.5\% , 0.4\%, 0.3\%, 0.2\% and 0.1\% 
Producers.}
\end{figure}
\item Profit for government\\[0.1cm]

Fig.1b shows the government profit as a function of the Tobin tax for the RM. 
We took again maxwin= 5\%.\\
 From 
top to bottom the graph shows the behaviour when we take the following 
values for the  Producers: 0.5 \% , 0.4 \% , 0.3 \% , 0.2 \%  and 
finally 0.1 \%.\\

In fig.2a we take 1\% Producers and fit the first part by the 
straight line\\
$ y=(3983\pm630)+(126550\pm14585)*x$ (where $x$ represents the Tobin tax rate 
and $y$ the profit for government)\\[0.1cm]
 and the last part by  the straight line\\
$ y=(17762\pm163)+(10622\pm356)*x.$\\[0.1cm]
 Both straight lines intersect at $(0.120\%\pm0.009\%)$ Tobin tax.\\[0.1cm]

 Thus if a Tobin tax should be introduced by government, we suggest a 
value of $\approx 0.12\%$. This is the case where government makes the highest 
profit from speculators, on condition of returns between $\approx -0.8\%$ 
and $\approx +0.8\%$ (fig.8b). These returns are 
reasonable because we have seen above, speculators try to take advantage 
of profits below the $10^{-3}$ border. If we take one time step as a week, 
the extreme return of 0.8\% corresponds to an interest calculation of 
83.2\% per annum. Furthermore we found in our calculations that the value of 
$\approx 0.12\%$ Tobin tax is independent of the number of Producers 
as long as there are not more than 5\% Producers. Notice that in reality 
there are far less Producers than 5\%, so we can propose a 
tax of $\approx0.12\%$. \\
The straight line with which we fitted the first part of the graph corresponds 
to ignoring the Tobin tax, in the way that is not dependent 
on the Producers. The second straight line fitted the part where 
only the Producers trade.
\begin{figure}[!]
\begin{center}
\includegraphics[angle=-90,scale=0.31]{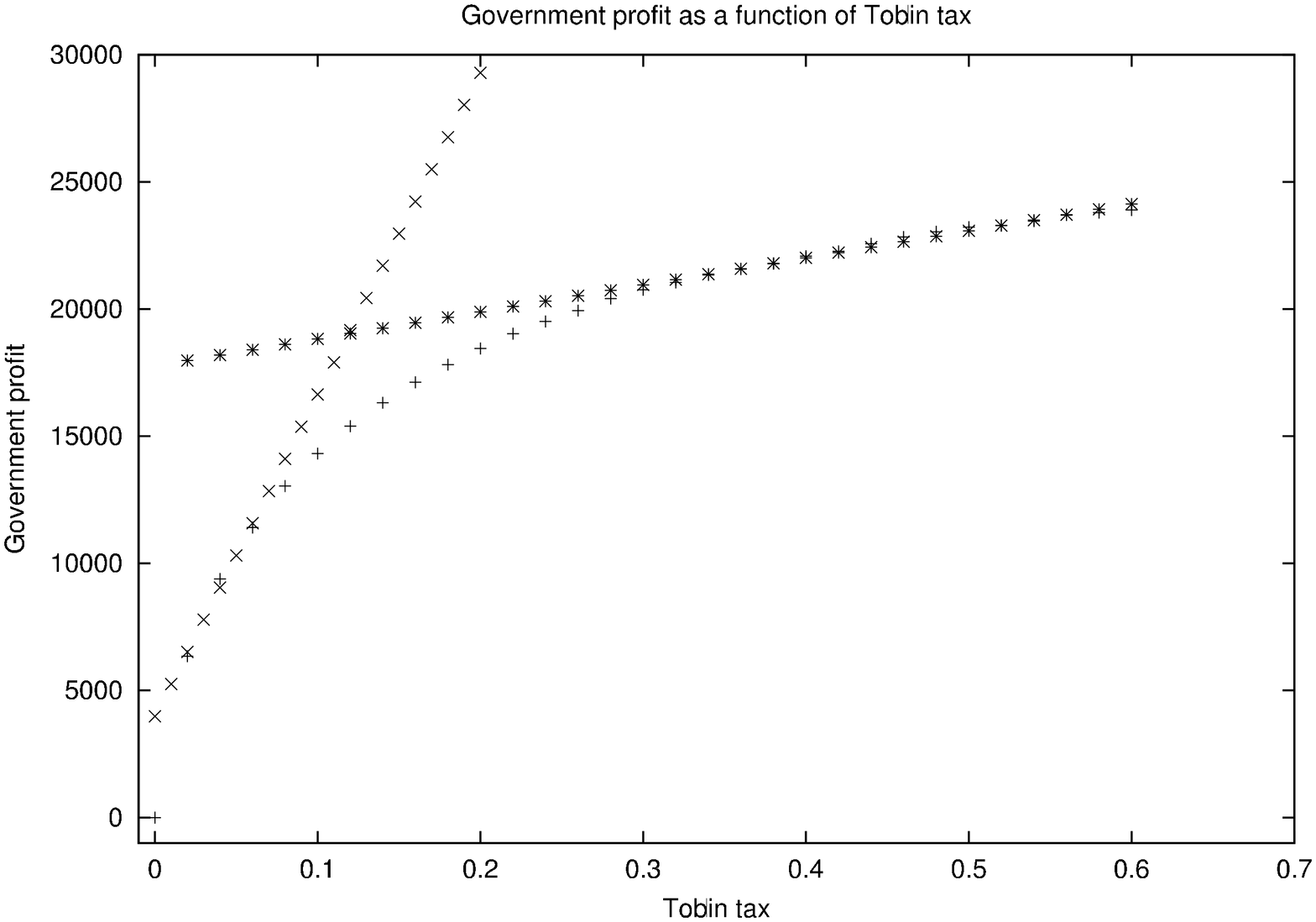}
\includegraphics[angle=-90,scale=0.31]{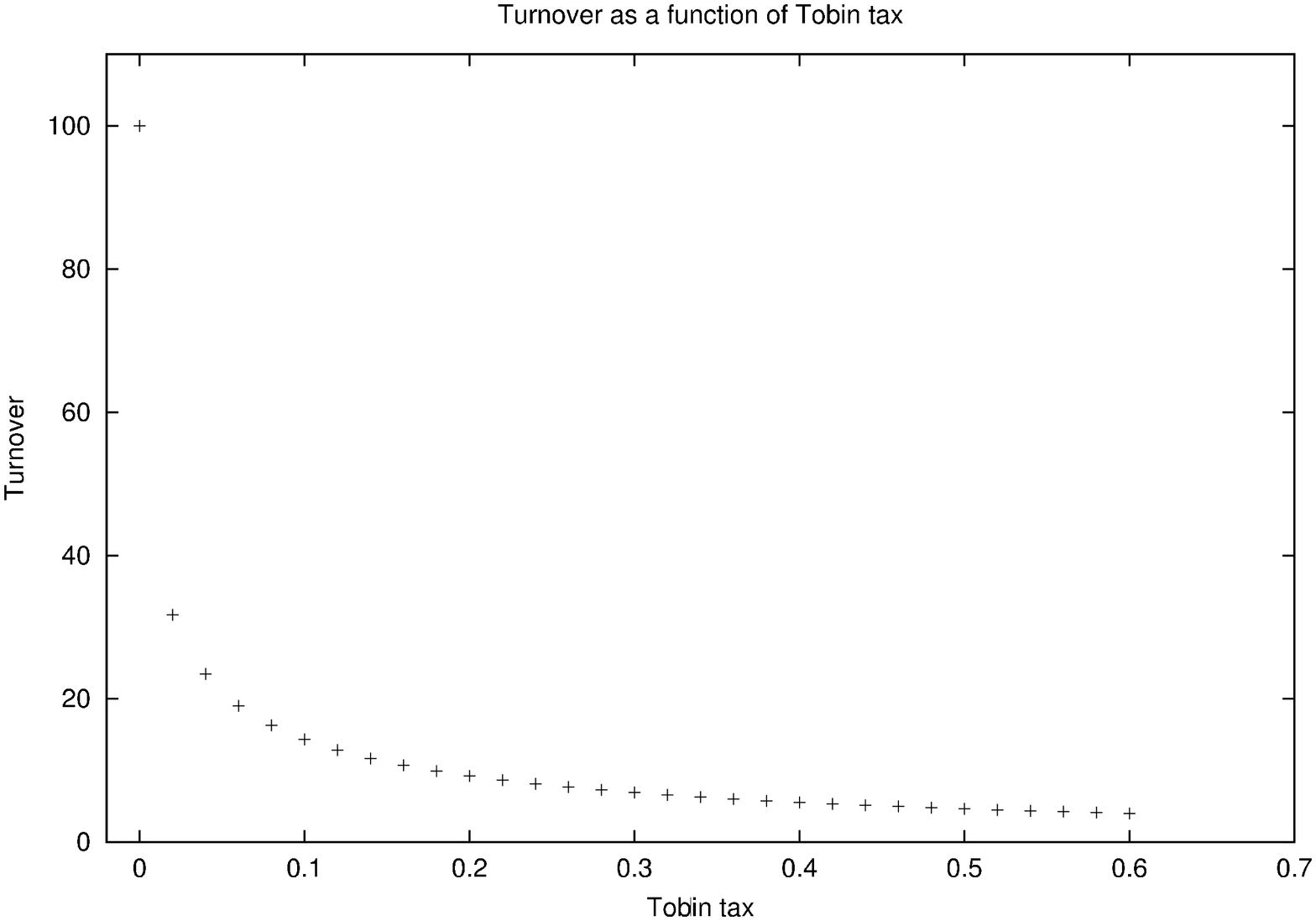}
\end{center}
\caption{In this figure we take 1\% Producers.
a) Profit for government as a function of Tobin tax. 
b) Turnover as a function of Tobin tax.} 
\end{figure}
\item Turnover as a function of Tobin tax\\[0.2cm]
In fig.2b we take 1\% Producers and maxwin= 5\%. We show the way
 the turnover will decrease if we introduce a Tobin tax. 
If we take the proposed value of 0.12\%, 
the turnover decreases by about 87\%. \\[0.1cm]

\item Profit for government as a function of Tobin tax. \\[0.1cm]
We will compare  RM and SCBM. \\[0.2cm]
\begin{itemize}
\item At first we take maxwin= 5\% \\[0.2cm]
From top to bottom fig.3a (RM) and fig.3b (SCBM) show the curves for 0.005\%, 0.004\%,
 0.0035\% and finally 0.003\% Producers.

\begin{figure}[!]
\begin{center}
\includegraphics[angle=-90,scale=0.31]{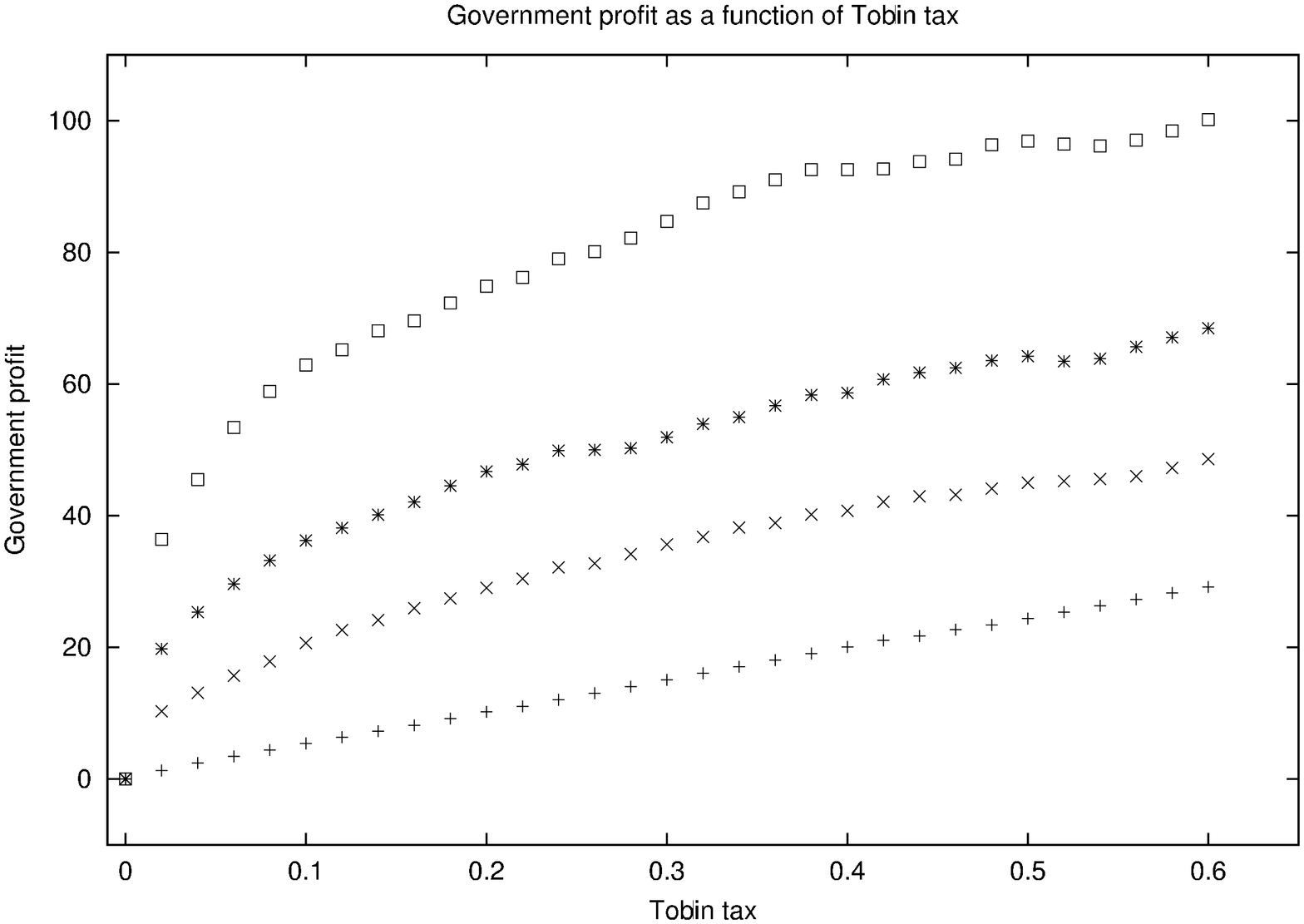}
\includegraphics[angle=-90,scale=0.31]{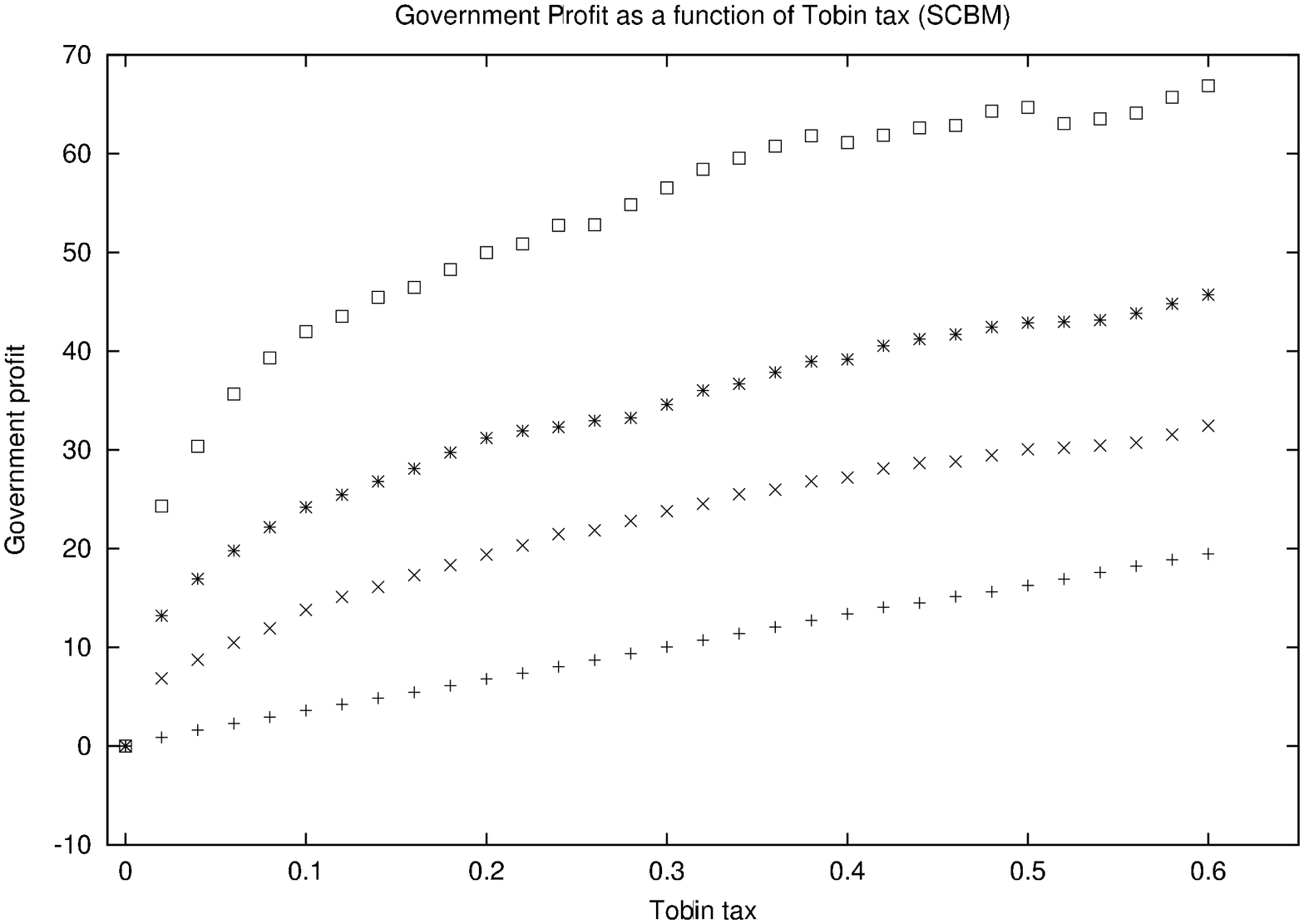}
\end{center}
\caption{Profit for government as a function of Tobin tax, we take from upper to bottom
 0.005\%, 0.004\%, 0.0035\%, 0.003\% Producers. a) RM. b) SCBM.}
\end{figure}

 The fig.4a (RM) and fig.4b (SCBM)
 show that there 
won't be a difference between the 
values one should propose for Tobin tax, when we compare both models.
 We take 0.005\% Producers here.\\
The first part of figure 4a is fitted by \\
$y=(28\pm1)+(425\pm17)*x$\\[0.1cm]
The second straight line which is used to fit the second part is:\\
$y=(71\pm2)+(50\pm4)$*x\\[0.2cm]
The first part of figure 4b is fitted by \\
$y=(19\pm1)+(281\pm11)*x$\\[0.1cm]
The second straight line which is used to fit the second part is:\\
$y=(46\pm1)+(35\pm3)*x$\\[0.1cm]

Again we can state that the convenient Tobin tax is $(0.113\%\pm0.013\%)$ 
(which corresponds to the value we got above within the error margin) when we 
relate to zero (RM).\\
And when we take the model similar to Cont-Bouchaud (SCBM), we get for government 
a convenient Tobin tax of $(0.118\%\pm0.008\%)$.
\begin{figure}[!]
\begin{center}
\includegraphics[angle=-90,scale=0.31]{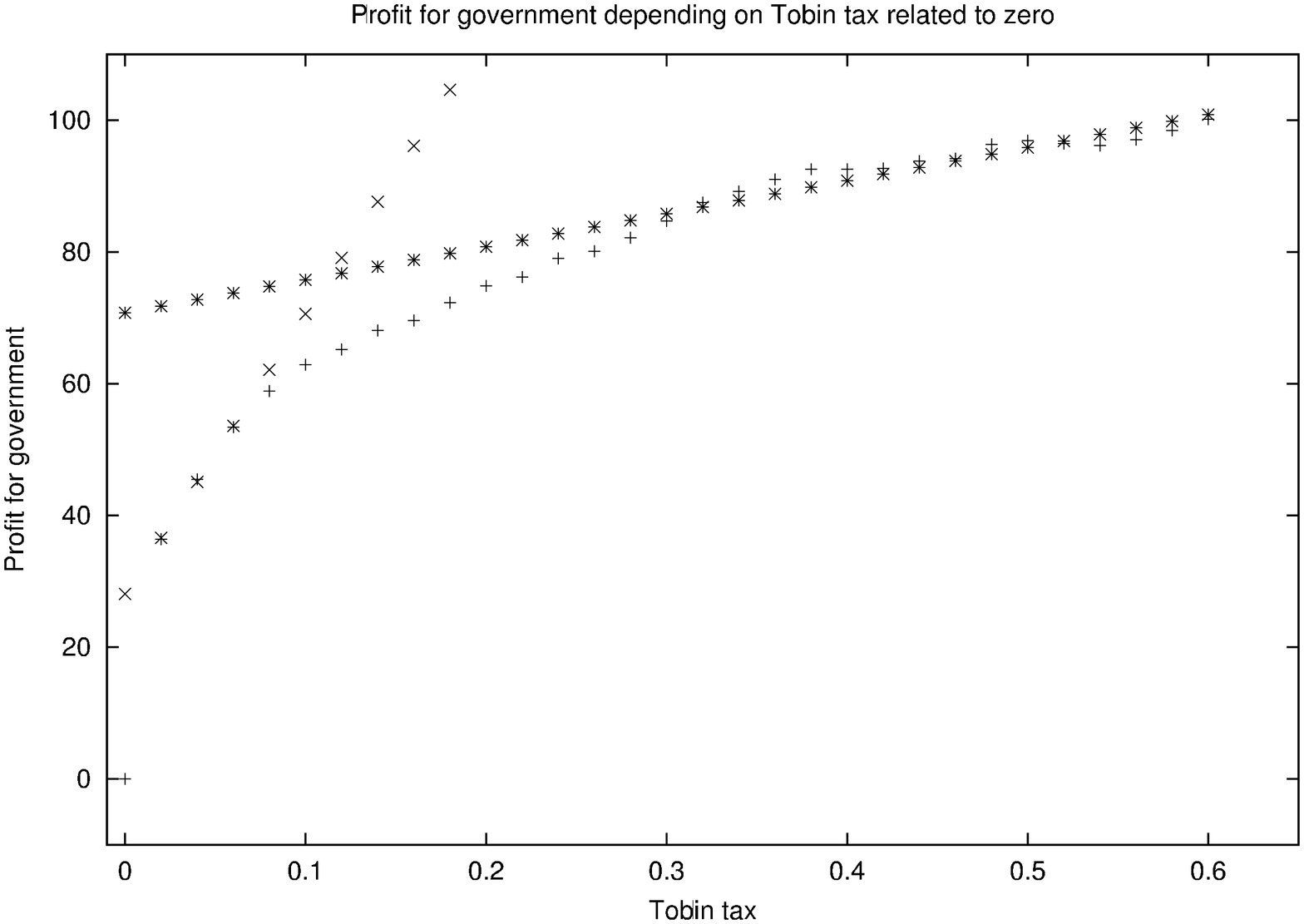}
\includegraphics[angle=-90,scale=0.31]{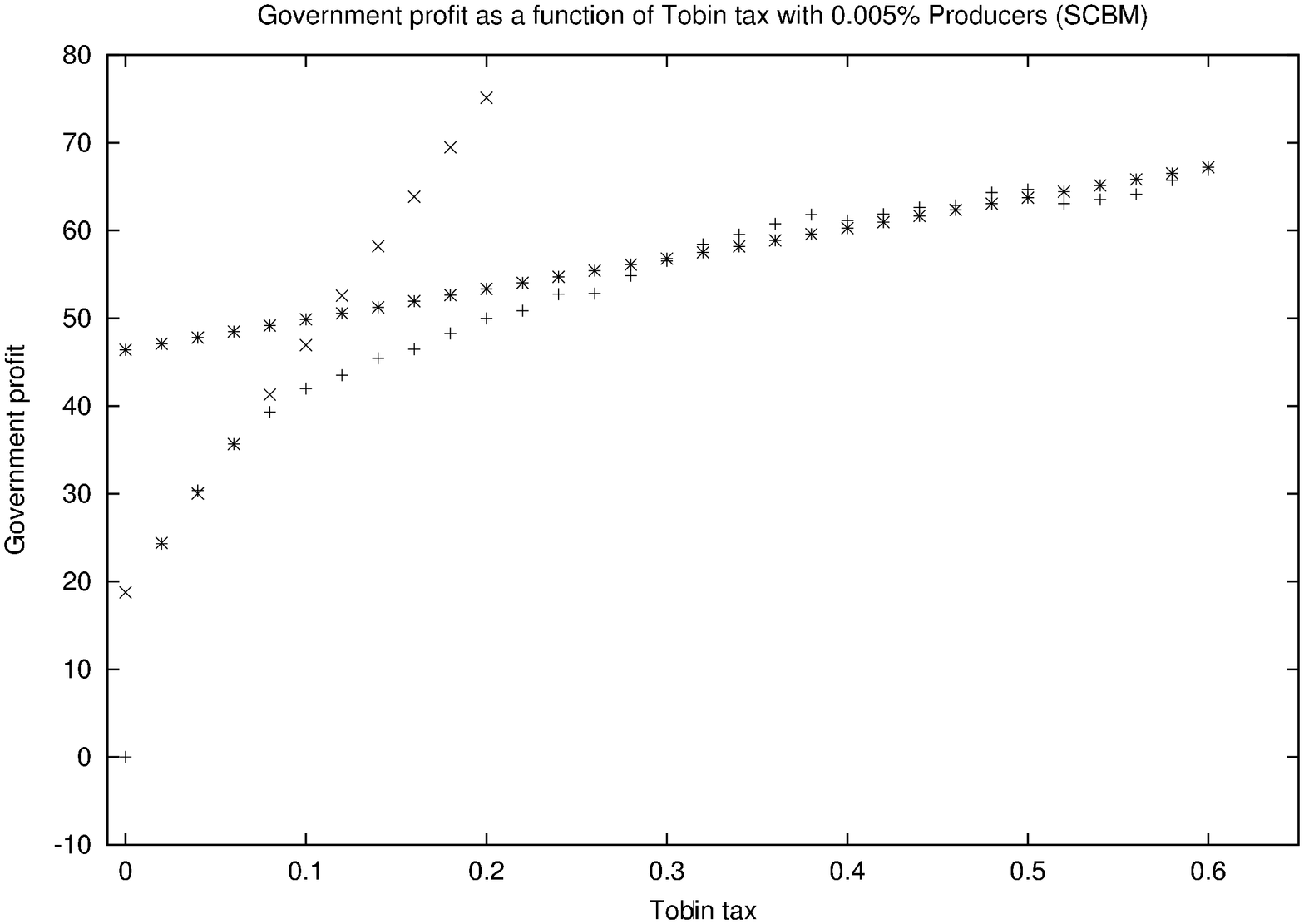}
\end{center}
\caption{Profit for government as a function of Tobin tax, we take 0.005\% Producers 
and maxwin= 5\%. a) RM. b) SCBM.
}
\end{figure}

\item Now we take maxwin=50\% \\[0.2cm]
The returns one gets now are actually between $-8\%$ and $+8\%$.
From top to bottom in fig.5a and fig.5b are  
curves for 0.005\%, 0.004\%, 0.0035\% and finally 0.003\% Producers.

\begin{figure}[!]
\begin{center}
\includegraphics[angle=-90,scale=0.31]{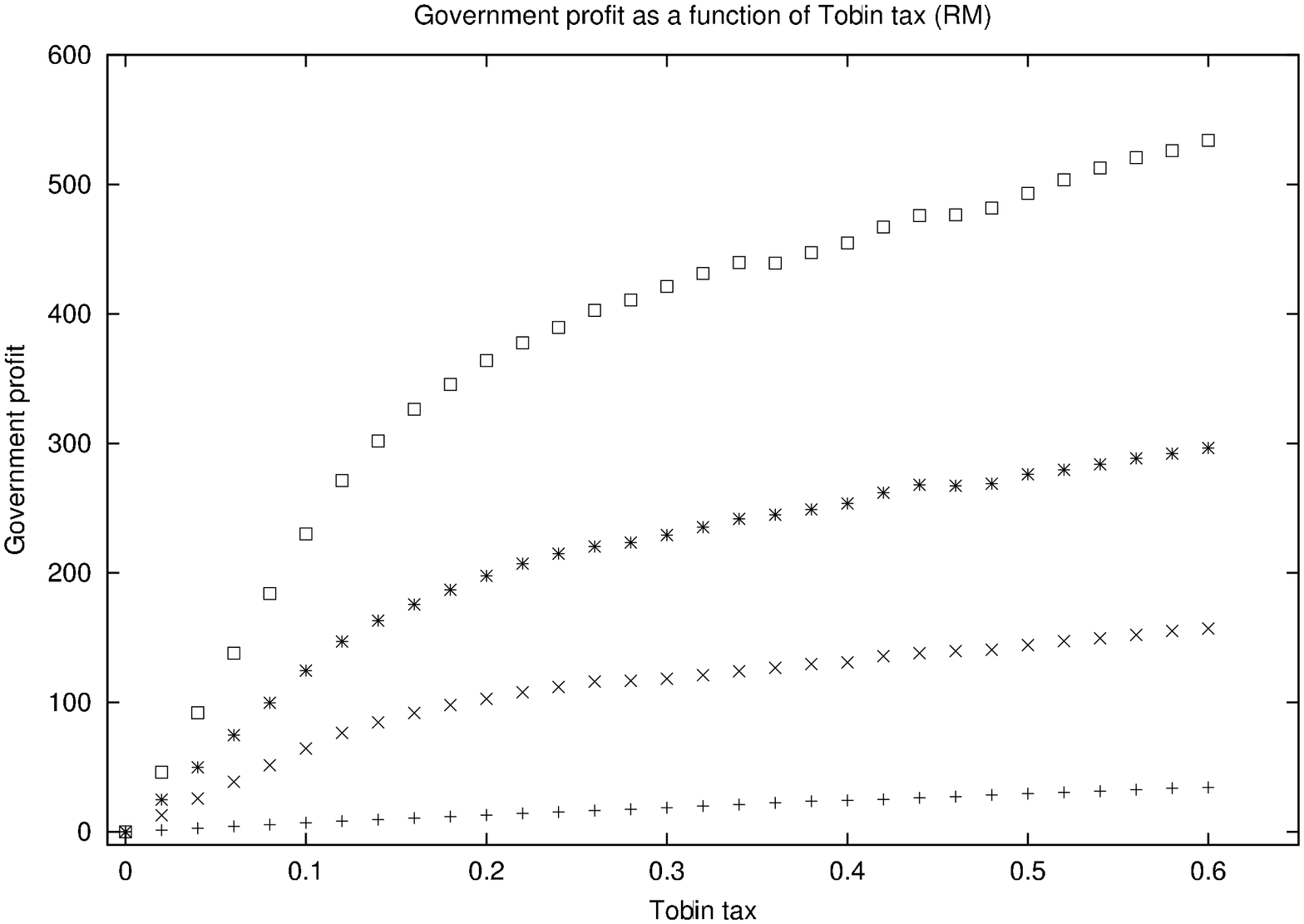}
\includegraphics[angle=-90,scale=0.31]{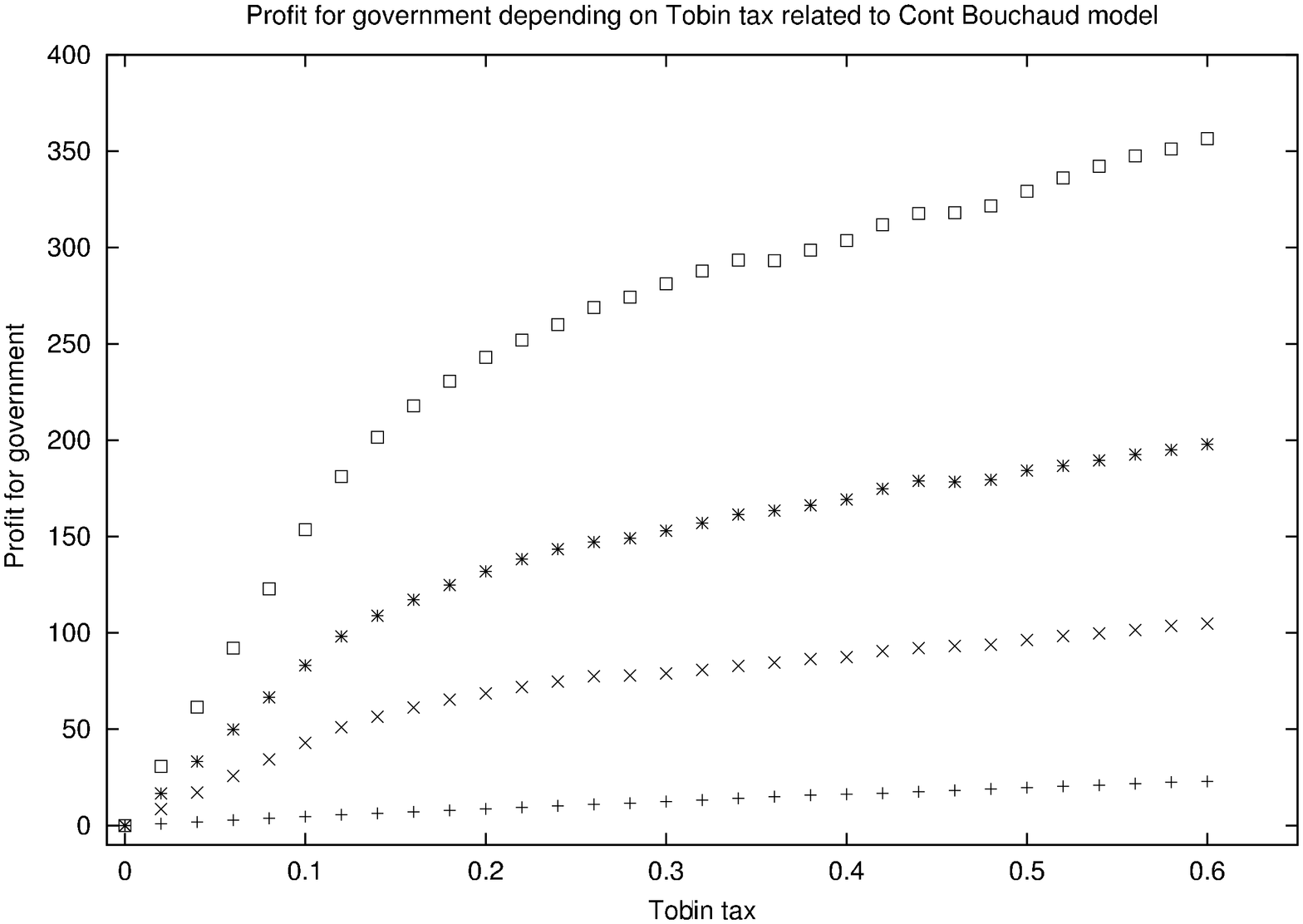}
\end{center}
\caption{Profit for government as a function of Tobin tax, we take 0.005\%, 0.004\%, 
0.0035\% and 0.003\% Producers and maxwin= 50\%. a) RM. 
b) SCBM.
}
\end{figure}

We see in fig.5a and fig.5b that there is an expected difference for proposing a 
Tobin tax depending on the value of maxwin.
\begin{figure}[!]
\begin{center}
\includegraphics[angle=-90,scale=0.31]{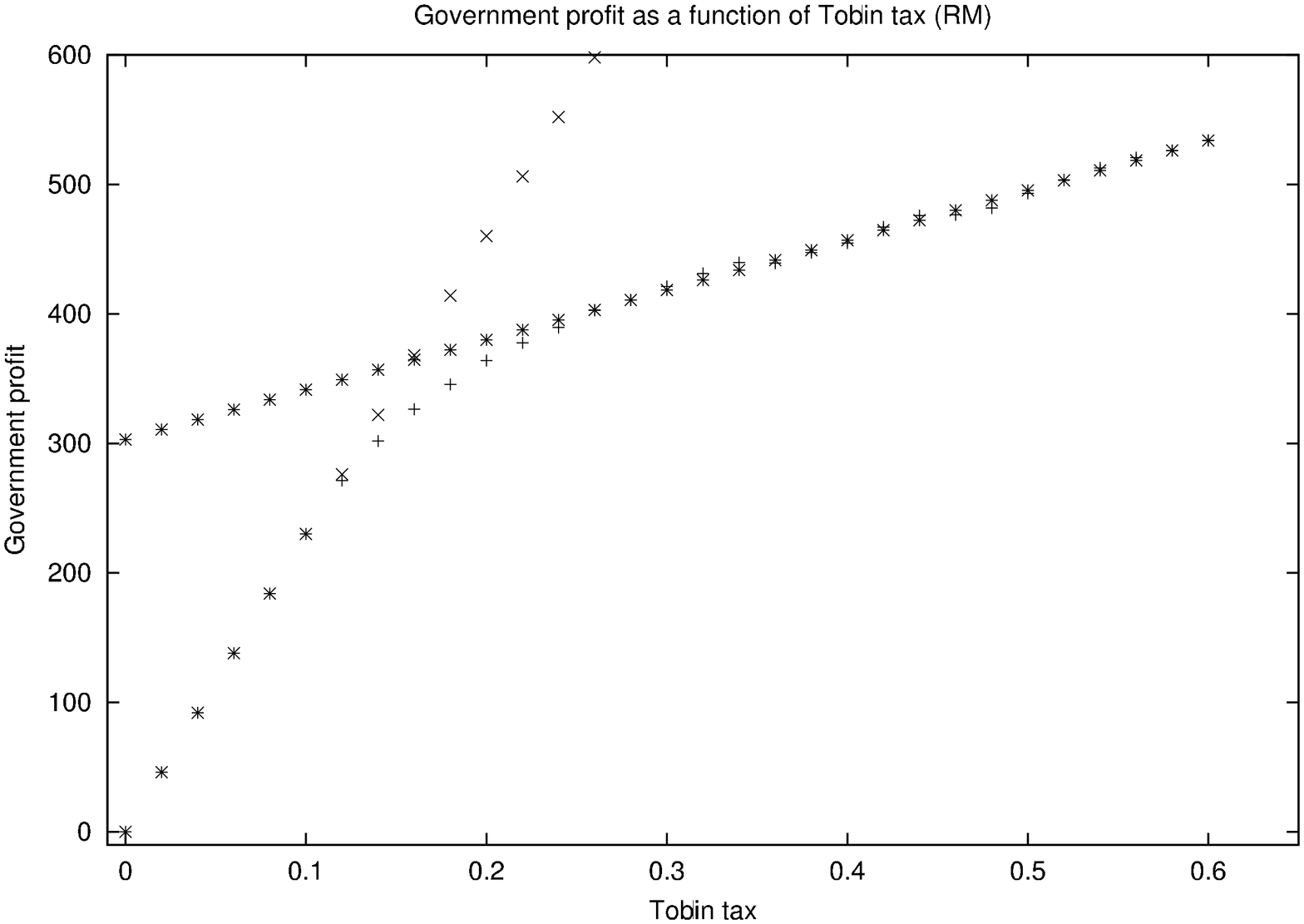}
\includegraphics[angle=-90,scale=0.31]{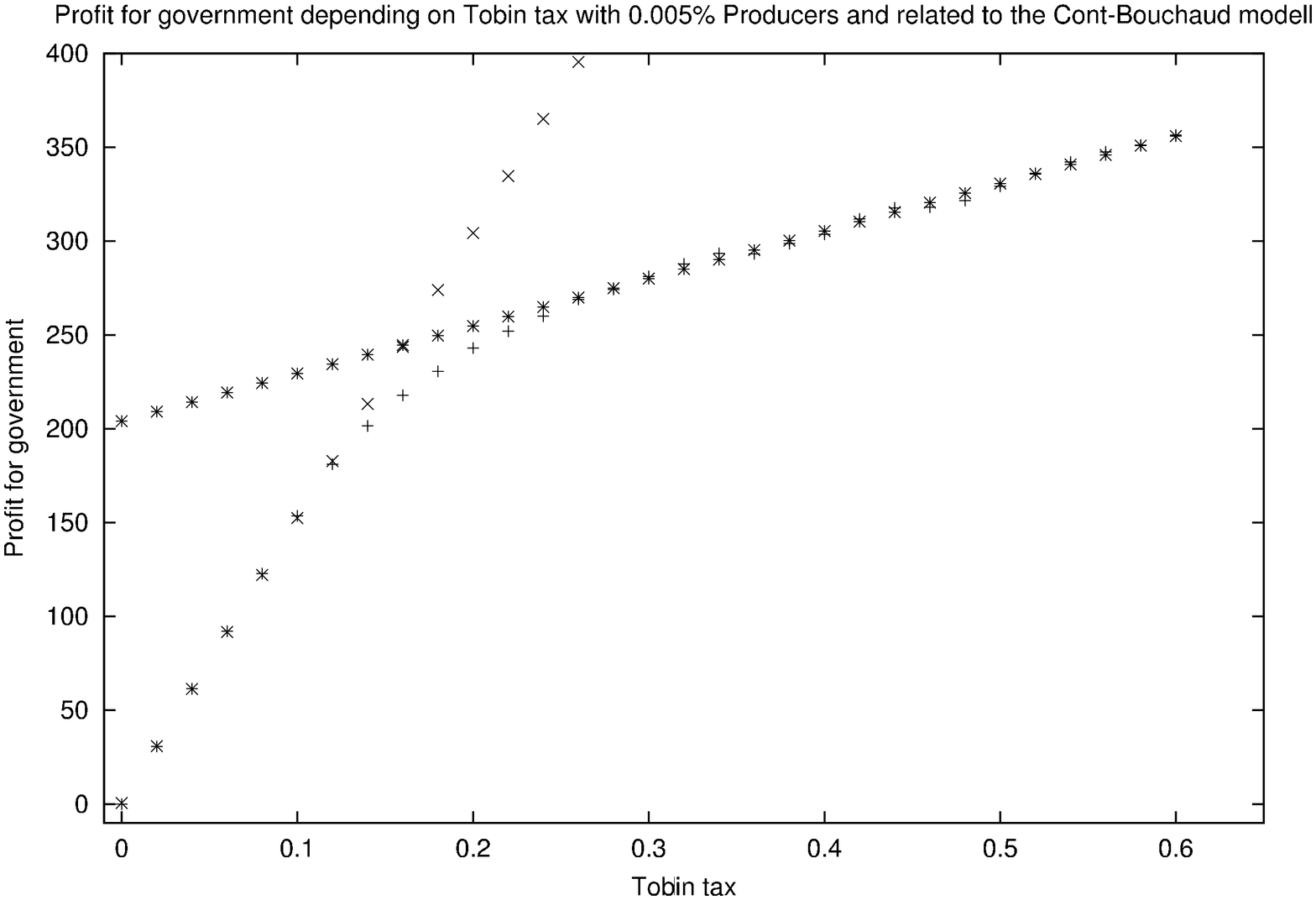}
\end{center}
\caption{Profit for government as a function of Tobin tax, we take 0.005\% Producers 
and maxwin=50\%. a) RM. b) SCBM.} 
\end{figure}

The first part of fig.6a is fitted by  \\
$y=(0\pm0)+(2301\pm0)*x$\\[0.1cm]
The second straight line which is used to fit the second part is:\\
$y=(303\pm3)+(385\pm7)$*x\\[0.1cm]
Both straight lines intersect at $(0.158\pm0.002)\%$ Tobin tax.\\[0.2cm]
The first part of fig. 6b is fitted by \\
$y=(0.55\pm0.69)+(1519\pm10)*x$\\[0.1cm]
The second straight line which is used to fit the second part is:\\
$y=(204\pm2)+(253\pm5)*x$\\[0.1cm]
Here both straight lines intersect at $(0.161\pm0.004)\%$ Tobin tax.\\[0.2cm]
Both values we get here for Tobin tax agree within the margin errors.\\
Result here: If we take maxwin as 50\% we should propose a Tobin tax 
of 0.16\%.

\end{itemize}

\item Profit for government as a function of Tobin tax with 
increasing Producers.\\[0.2cm]
\begin{figure}[!]
\begin{center}
\includegraphics[angle=-90,scale=0.31]{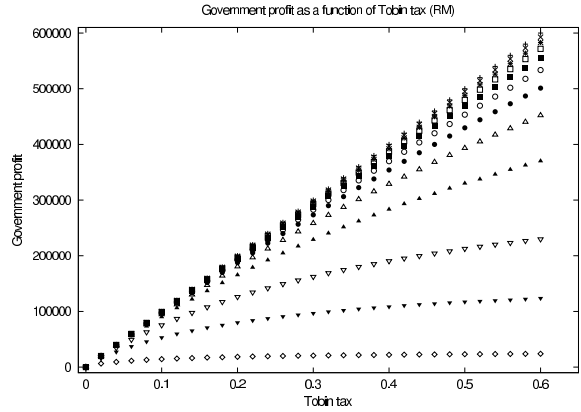}
\end{center}
\caption{Profit for government as a function of Tobin tax with increasing 
Producers,maxwin=5\%}
\end{figure}

Here in fig.7 we have from top to bottom 100\%, 90\%, 80\%, 70\%, 60\%, 50\%, 40\%, 
30\%, 20\%, 10\%, 5\% and finally 1\%.
One can see that we get  straight lines as long as there are 50\% Producers . 
These straight lines show that there is an insensitivity against the Tobin tax.\\
But in real currency markets there are not so many Producers. The number of 
Producers will converge to a very small number. But it does not converge 
to zero because there are people who  must sell because they need money.\\
We get the following straight lines from top to bottom:\\[0.2cm]
\begin{tabular}{|l|l|}
\hline
\multicolumn{2}{|c|}{
Straight lines which fit profit as a function of Tobin tax }\\
\hline
Producers            &Straight line\\
\hline
100\% $\Leftrightarrow$ Cont-Bouchaud        &$y=0$\\
\hline                       
90\%         &$y=(987570\pm607)*x-(1168\pm212)$\\
\hline
80\%       &$y=(971500\pm1335)*x-(2615\pm466)$ \\
\hline
70\%       &$y=(952320\pm2297)*x-(4493\pm802)$\\
\hline
60\%         &$y=(924940\pm3561)*x-(6987\pm1243)$\\
\hline
50\%       &$y=(888740\pm5244)*x-(10369\pm1832)$ \\
\hline
70\%       &$y=(833460\pm7571)*x-(15207\pm2644)$\\

\hline
\end{tabular}  \\[0.2cm]

\item Return as a function of time\\[0.2cm]
In the following Figs.8 and 9 the ``+'' sign represents the returns in 
the original Cont-Bouchaud model, the ``x'' sign represents the returns when $r_-$ 
should be greater or equal to zero, `` * '' shows the 
behaviour when a Tobin tax of 0.01\% is introduced and the empty squares 
stand for returns when a tax of 0.16\% is introduced.\\
We took a square lattice of length 31 and an average about 100 
cluster configurations.\\
In fig.8a (maxwin= 50\%) and fig.8b (maxwin= 5\%) we took 1\% Producers. 
In fig.9 we took only 0.005\% Producers (maxwin= 5\%).\\
\begin{figure}[!]
\begin{center}
\includegraphics[angle=-90,scale=0.31]{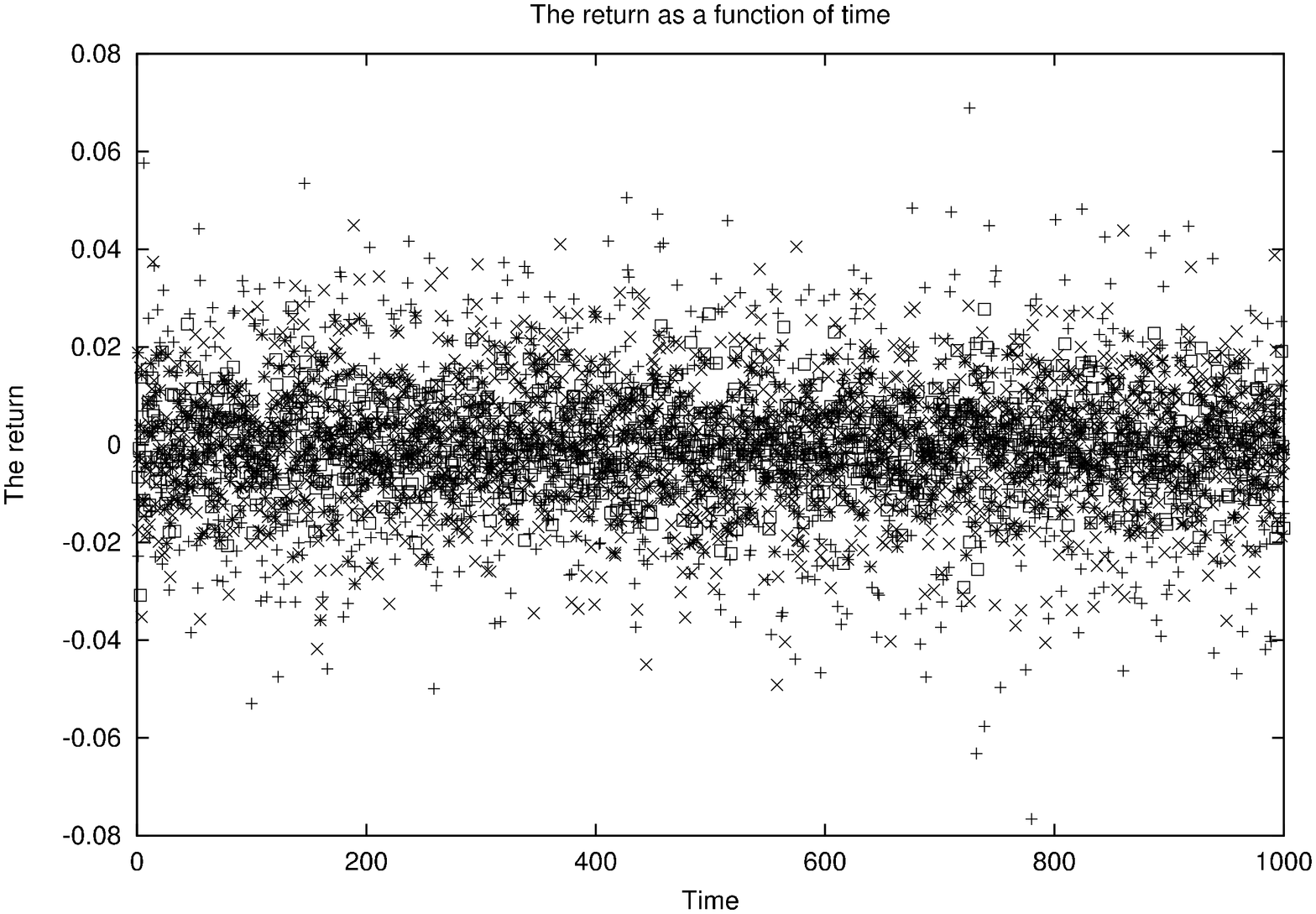}
\includegraphics[angle=-90,scale=0.31]{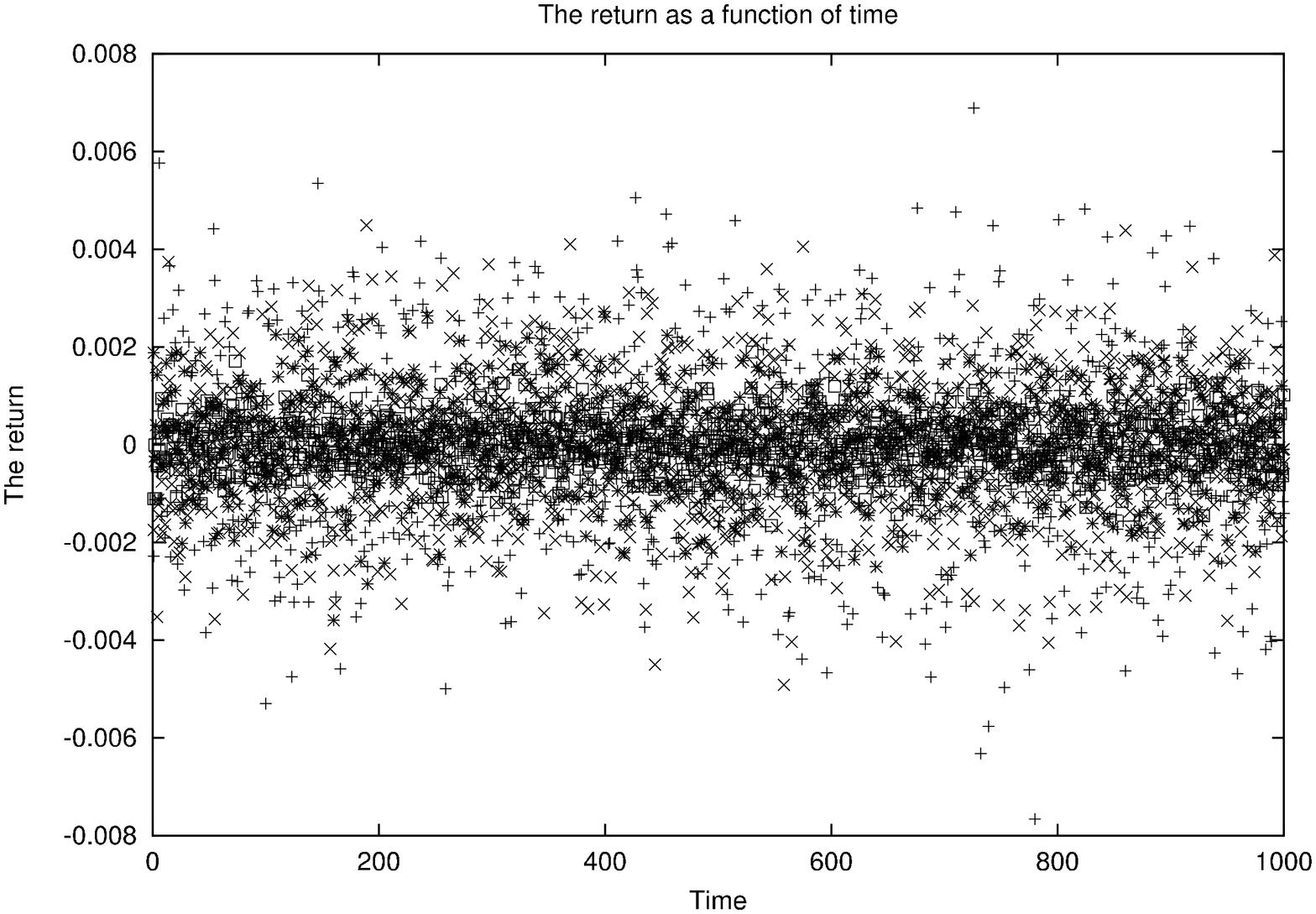}
\end{center}
\caption{Return as a function of time, 1\% Producers. a) maxwin=50\%, b) maxwin=5\%}
\end{figure}

\begin{figure}[!]
\begin{center}
\includegraphics[angle=-90,scale=0.31]{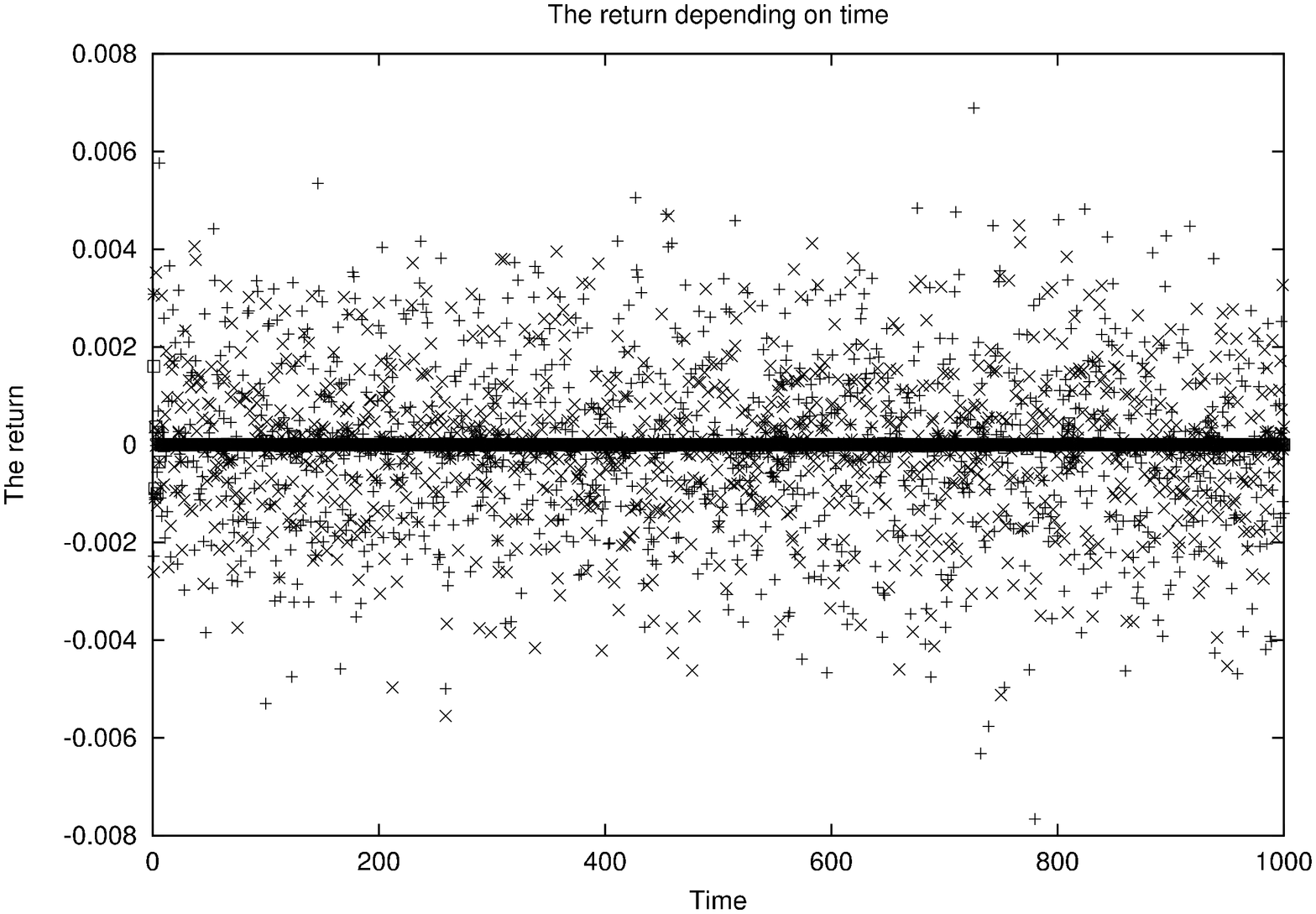}
\end{center}
\caption{Return as a function of time, maxwin=5\%, Producers= 0.005\%}
\end{figure}

\item The Return Histogram\\[0.2cm]
The deviation of the heavy tails from the Gaussian will be 
characterized by a significant excess kurtosis which is defined by 

\begin{equation}
\label{1}
\kappa=\frac{{\mu_4}}{{\sigma^4}}-3 \quad ,
\end{equation}

where $\mu_4$ is the fourth central moment and $\sigma$ the standard 
deviation of the returns. $\kappa$ should be zero for a normal distribution 
but it ranges between 2 and 50 for daily returns and is even higher for 
intra day data.
The fat tails correspond to large fluctuations in returns.\\
For the following calculations we took a square lattice of length 101. 
We took an average of about 100 cluster configurations and made 100000 iterations.\\
Here are the prices not normalized and we take 1\% Producers.\\ 
Figure 17 shows from outside to inside the Gaussian for the original Cont-Bouchaud 
model for which we get a kurtosis of $\kappa=0.58$ here (probably this is due to the 
small lattice). The next interior curve represents the case when people act when $r_-$ 
is greater or equal zero ($\kappa=3.71$). What follows is the curve where Tobin tax 
amounts 0.01\% ($\kappa=8.38$) and for the last one a Tobin tax of 0.16\% ($\kappa=56.76$)
 is assumed.
In figure 16 we have add the following two cases: The Cont-Bouchaud model with activity 
a=10\% and a=1\%. The outer one of the two added curves corresponds to activity a=10\% 
($\kappa=4.96$) and the other one to a=1\% ($\kappa=54.72$).\\
From the picture we get the results that the effect of ``convergence'' to zero is the 
greater the greater Tobin tax is and the ``convergence'' to zero is for a $\rightarrow$ 
1\% not so drastic as for Tobin tax $\rightarrow$ 0.16\%.

\begin{figure}[!]
\begin{center}
\includegraphics[angle=-90,scale=0.5]{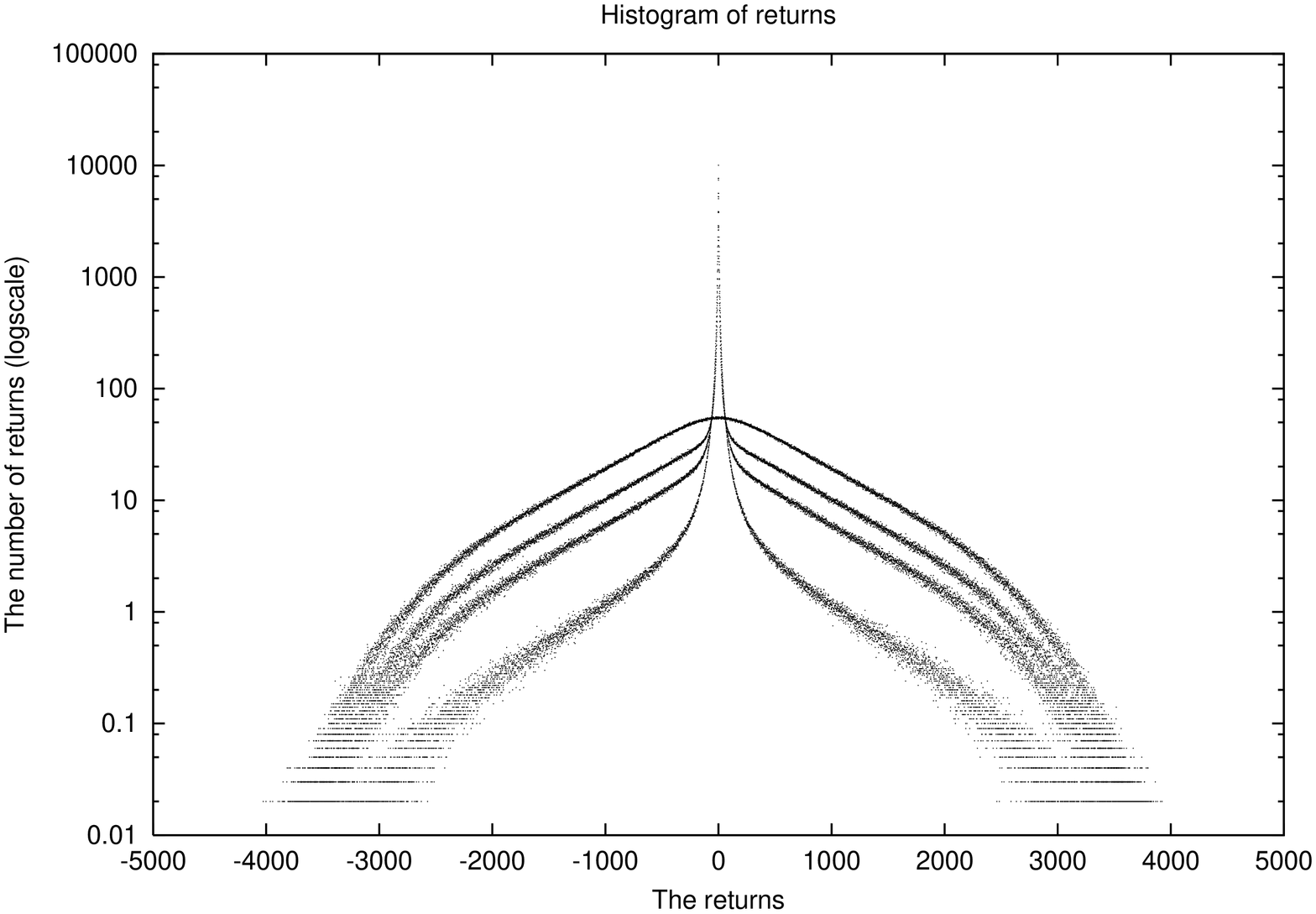}
\includegraphics[angle=-90,scale=0.5]{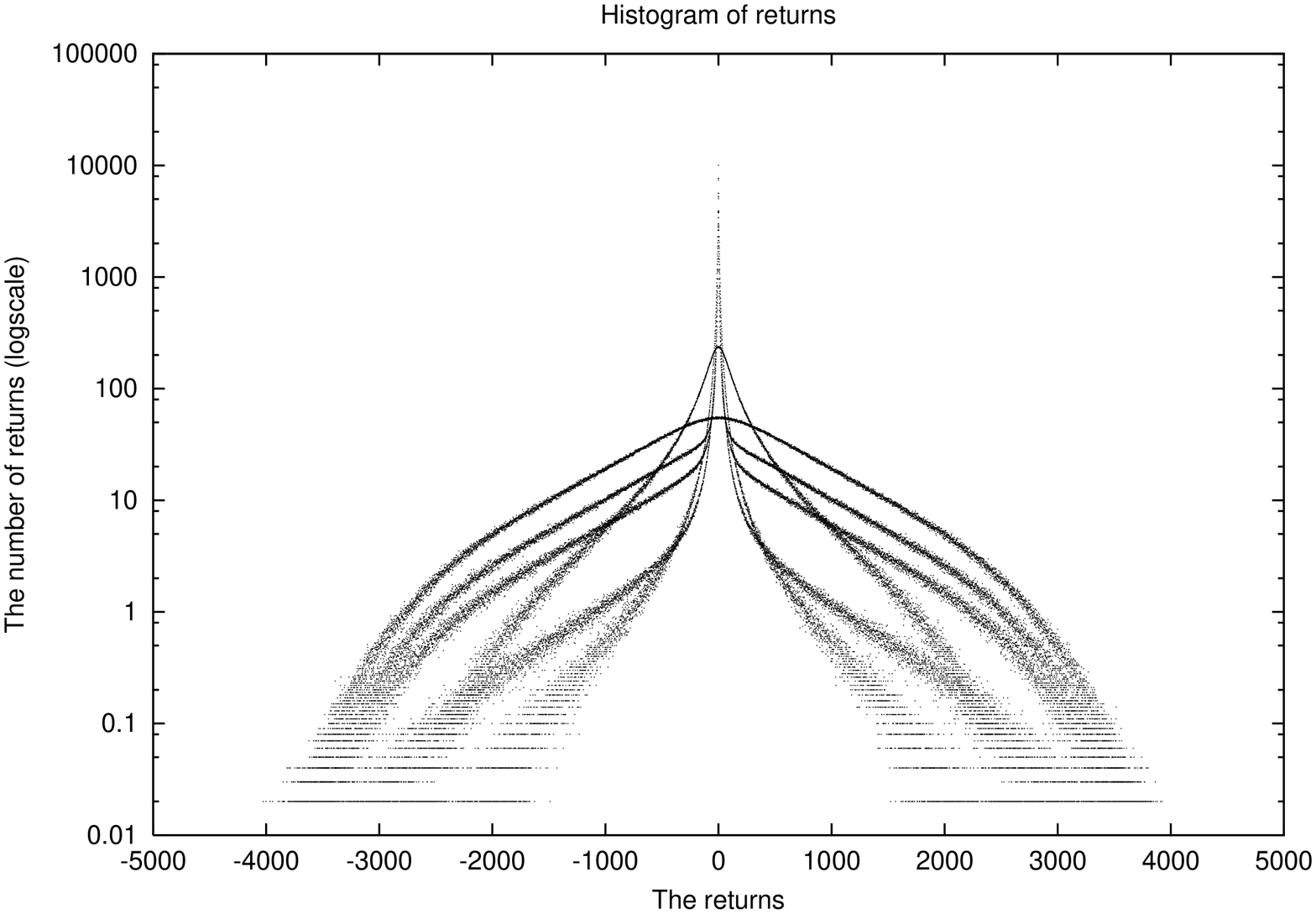}
\end{center}
\caption{a)Histogram of returns. b) Histogram of returns compared with different activities.}
\end{figure}

\item Number of speculating and trading investors\\[0.2cm]
\begin{figure}[!]
\begin{center}
\includegraphics[angle=-90,scale=0.31]{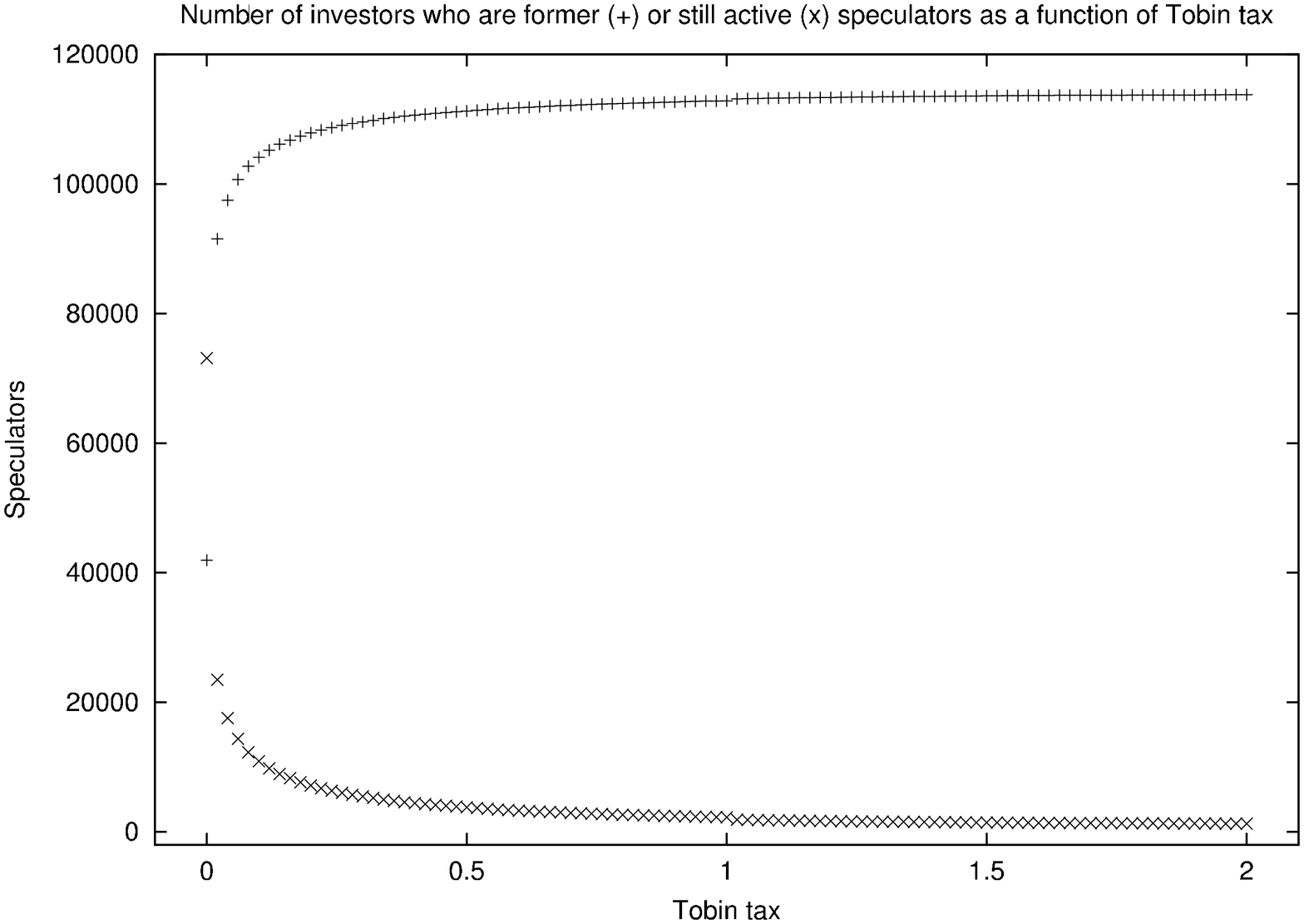}
\includegraphics[angle=-90,scale=0.31]{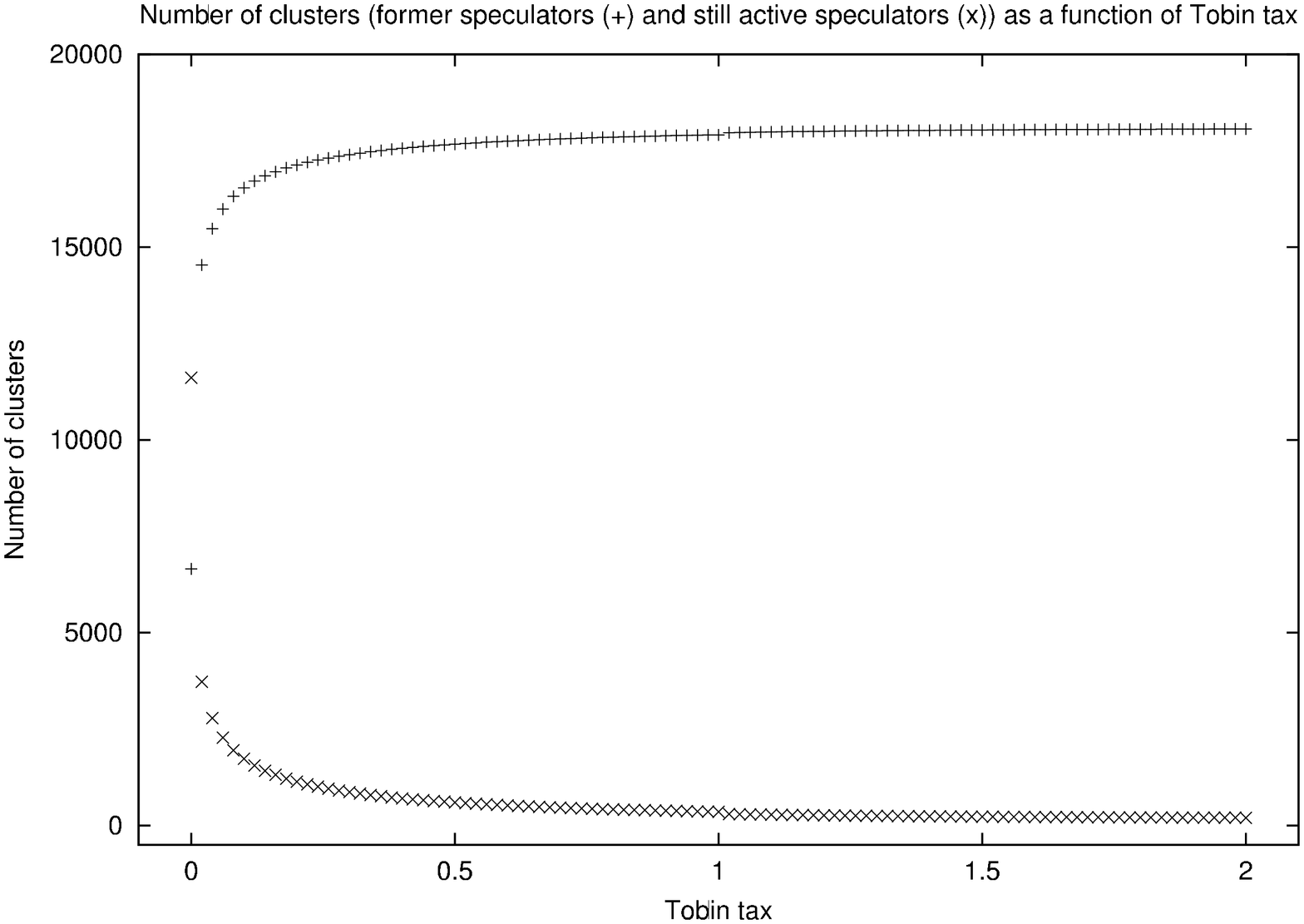}
\includegraphics[angle=-90,scale=0.31]{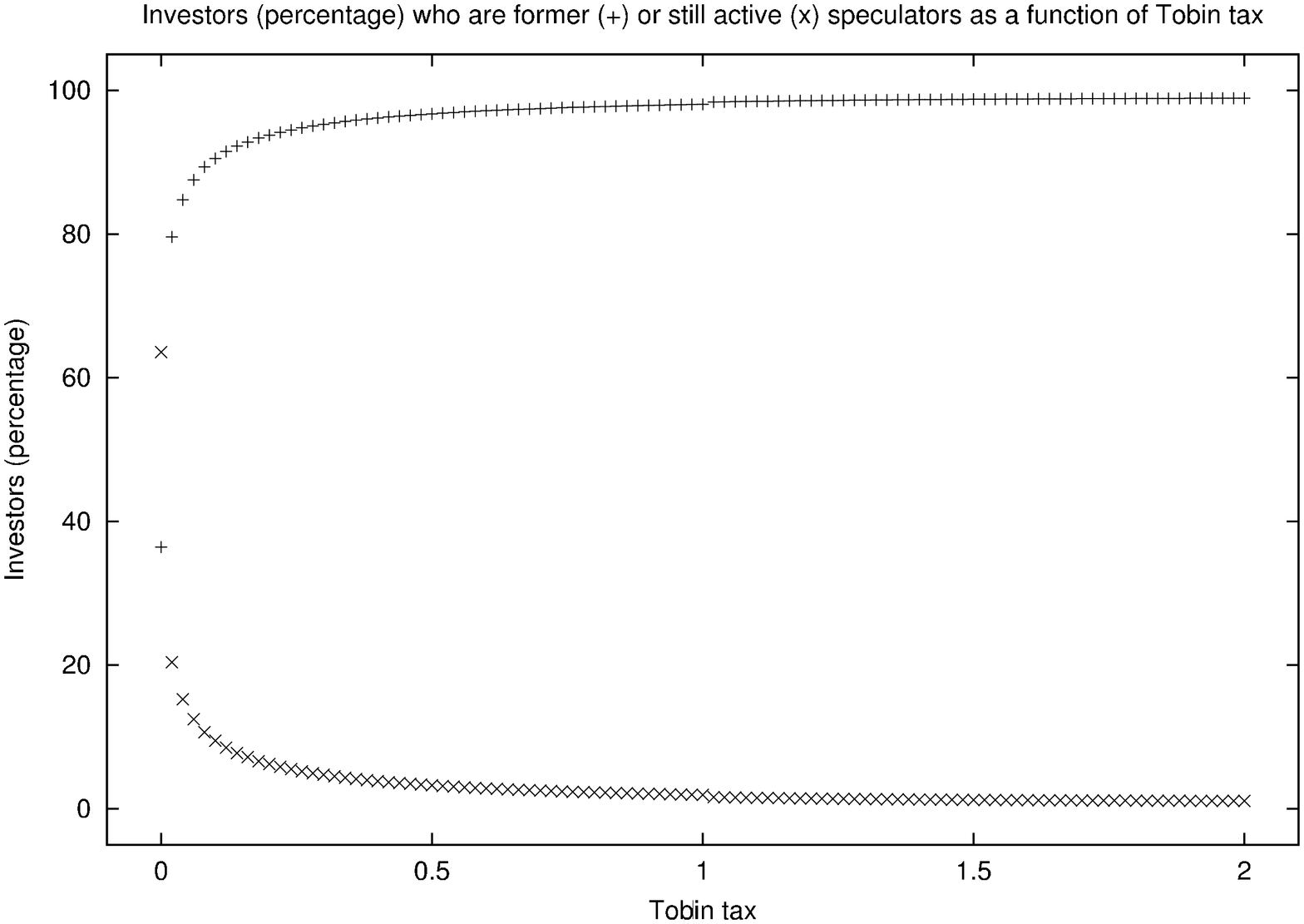}
\end{center}
\caption{RM. a) Number of former speculators (+) and still active speculators (x). 
b) Number of clusters (former speculators (+) and still active speculators (x). 
c) Former speculators (+) and still active speculators (x) (percentage).}
\end{figure}

For Fig.11, we sum up the number of investors who are former speculators (+) or still 
active speculators (x) over about 500 time steps for different values of Tobin tax rate. 
We define former speculators in the following way: The former speculators are those 
people who don't trade because they do not fulfill the condition $r_-$ greater than 
Tobin tax. Here is maxwin= 5\% and Producers= 1\%. Fig.11 shows the result for the RM. 
We have seen above that a Tobin tax rate of 0.12\% is convenient for government. 
But you can see that we have here 91\% former speculators and only 9\% still active 
speculators.\\
The SCBM gives nearly the same result except the fact that when Tobin tax is equal zero 
there are no former speculators and all available investors are active speculators.\\
The sum about number of clusters corresponding to the sum about the number of investors 
lies between 18265 and 18275.\\ 
If we take a look on the number of investors who are former or active speculators as a 
function of time, we get straight lines.\\[0.2cm]
\begin{tabular}{|l|l|}
\hline
\multicolumn{2}{|c|}{
Straight lines which fit the number of former speculators as a function of time}\\
\hline
Condition to trade   &Straight line\\
\hline
Cont-Bouchaud        &$y=0$\\
\hline                       
$r_-\ge0\%$          &$y=(83.833\pm0.002)*x-(66.062\pm0.466)$\\
\hline
$r_-\ge0.01\%$       &$y=(175.362\pm0.003)*x-(268.98\pm0.870)$ \\
\hline
$r_-\ge0.16\%$       &$y=(208.527\pm0.005)*x-(379.22\pm1.587)$\\
\hline
\end{tabular}  \\[0.2cm]
\begin{tabular}{|l|l|}
\hline
\multicolumn{2}{|c|}{
Straight lines which fit the number of still active speculators as a function of time}\\
\hline
Condition to trade   &Straight line\\
\hline
Cont-Bouchaud        &$y=(229.642\pm0.001)+(3.305\pm0.429)$\\
\hline                       
$r_-\ge0\%$          &$y=(145.774\pm0.003)*x+(68.755\pm0.764)$\\
\hline
$r_-\ge0.01\%$       &$y=(54.265\pm0.003)*x+(266.055\pm0.777)$ \\
\hline
$r_-\ge0.16\%$       &$y=(21.029\pm0.003)*x-(359.080\pm0.926)$\\
\hline
\end{tabular}  \\[0.2cm]

For every value of Tobin tax we have an average cluster size of $(6.42\pm0.02)$  and 
the ``infinite'' cluster we ignored because it causes only crashes and bubbles has 
an average size of $(311.38\pm0.97)$    .

\end{itemize}

\subsection{Results of the second modification of the Cont-Bouchaud model}
With this modification we did in generally the same calculations 
as described above but only for RM.

\begin{itemize}
\item Turnover as a function of time\\[0.1cm]
In fig.12a we see that trade will go to zero when there are no Producers.
We also see the way the trade volume will decrease with increasing 
Tobin tax when there are 0.5\% Producers. We took maxwin= 5\%\\[0.1cm] 
The graph shows from top to bottom:\\

The first line shows the turnover we get with $0\%$ Tobin tax and $0.5\%$ Producers. 
The next line is a turnover with  $0\%$ Tobin tax and $0\%$ Producers.\\
 We get the next following two lines which does not present a decrease of turnover 
after by about 100 time steps with $0.5\%$ and $1\%$ Tobin tax and in both 
cases $0.5\%$ Producers. The next two lines describe the behaviour for $0\%$ 
Producers for $0.5\%$ Tobin tax and $1\%$ Tobin tax.
What we see if there are no Producers, the turnover approaches 
zero the faster, the higher Tobin tax is, but the decay is not so drastic as 
we have seen with the first modification of the Cont-Bouchaud model.

\begin{figure}[!]
\begin{center}
\includegraphics[angle=-90,scale=0.5]{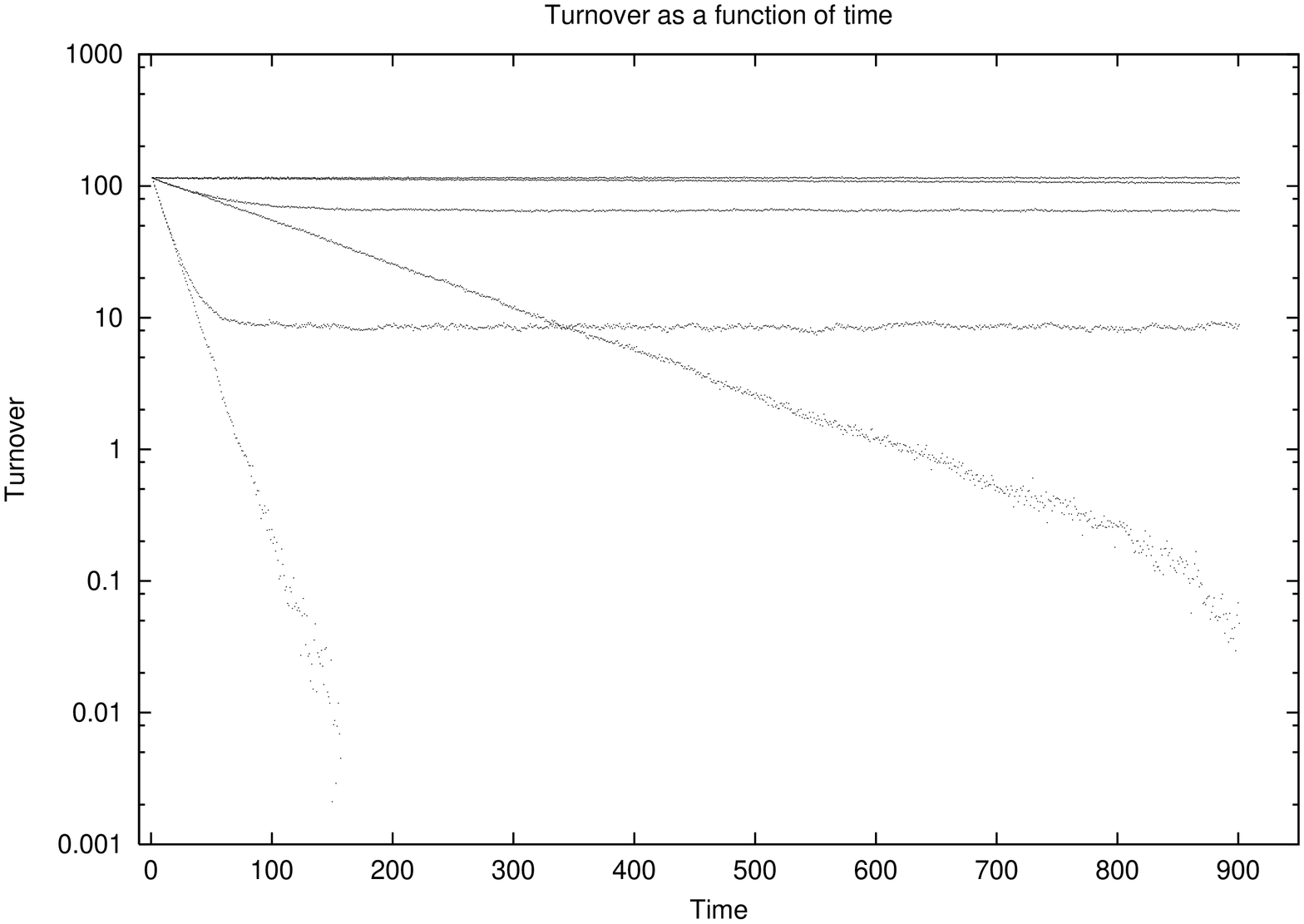}
\includegraphics[angle=-90,scale=0.5]{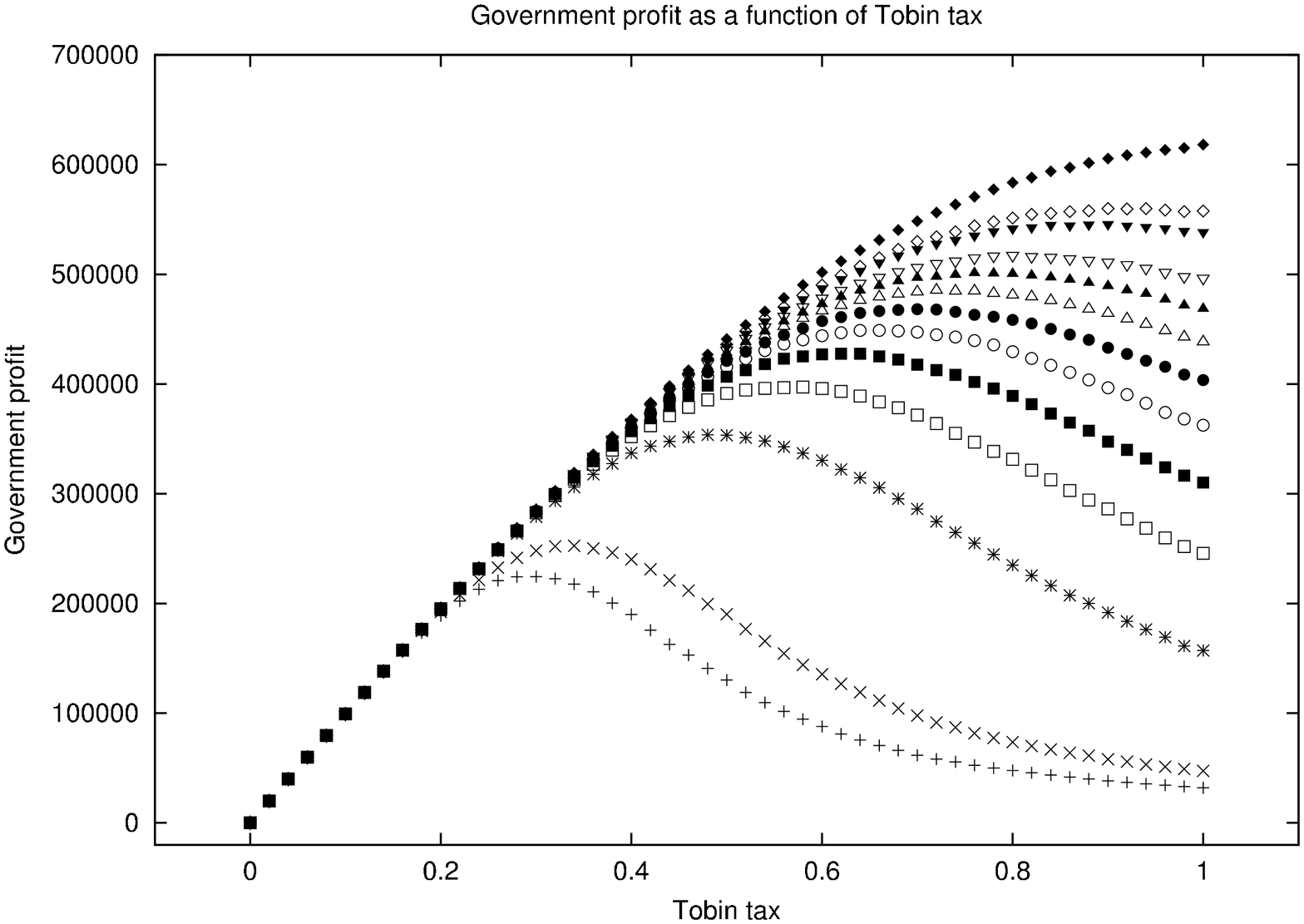}
\end{center}
\caption{
a) Turnover as a function of time. b) Profit for government as a function of 
Tobin tax. From top to bottom: 15\%, 11\%, 10\%, 8\%, 7\%, 6\%, 5\%, 4\%, 3\%, 
2\%, 1\%, 0.1\% and finally 0\% 
Producers.}
\end{figure}

\item Profit for government\\[0.1cm]
Fig.12b shows the profit for government as a function of the Tobin tax. From 
top to bottom the graph shows the behaviour when we take the following 
values for the  Producers:  15\%, 11\%, 10\%, 8\%, 7\%, 6\%, 5\%, 4\%, 3\%, 
2\%, 1\%, 0.1\% and finally 0\%.\\
We see here that the maximum goes to higher Tobin tax rates for an increasing 
number of Producers. We took here maxwin= 5\%. When we took maxwin= 50\% the 
maximum of profit for government is above 1.5\% Tobin tax.

\item Tobin tax to get maximal profit\\[0.1cm]
Fig.13a shows the  Tobin tax value one should take to get maximal income for 
government (Ttp) as a function of the Producers (P). The curve (+) is especially 
well fitted between 0\% and 2\% Producers (this corresponds to reality values 
for Producers) by $$ Ttp= \sqrt{\frac{P}{25}}+\frac{3}{10}$$.
\begin{figure}[!]
\begin{center}
\includegraphics[angle=-90,scale=0.31]{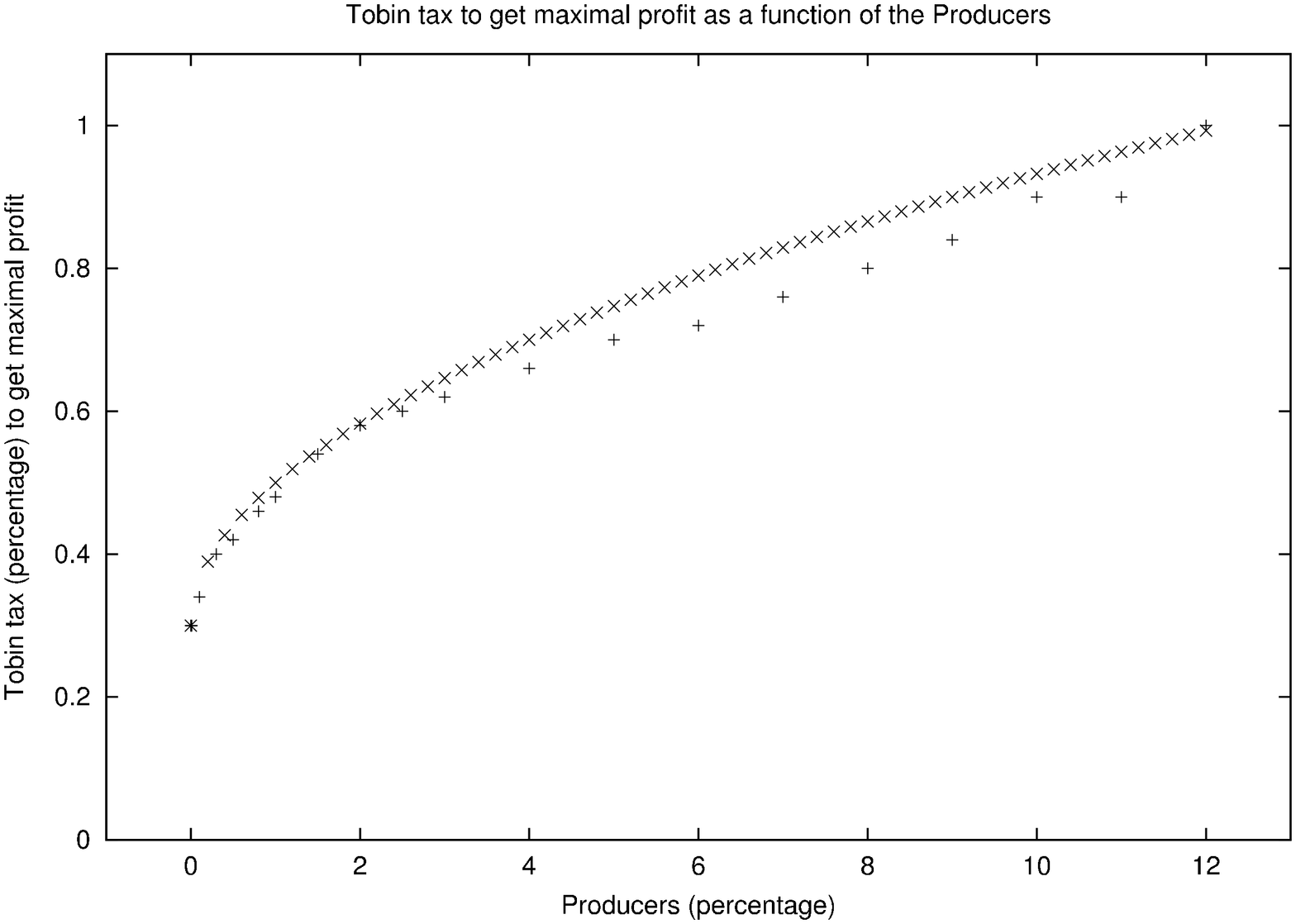}
\includegraphics[angle=-90,scale=0.31]{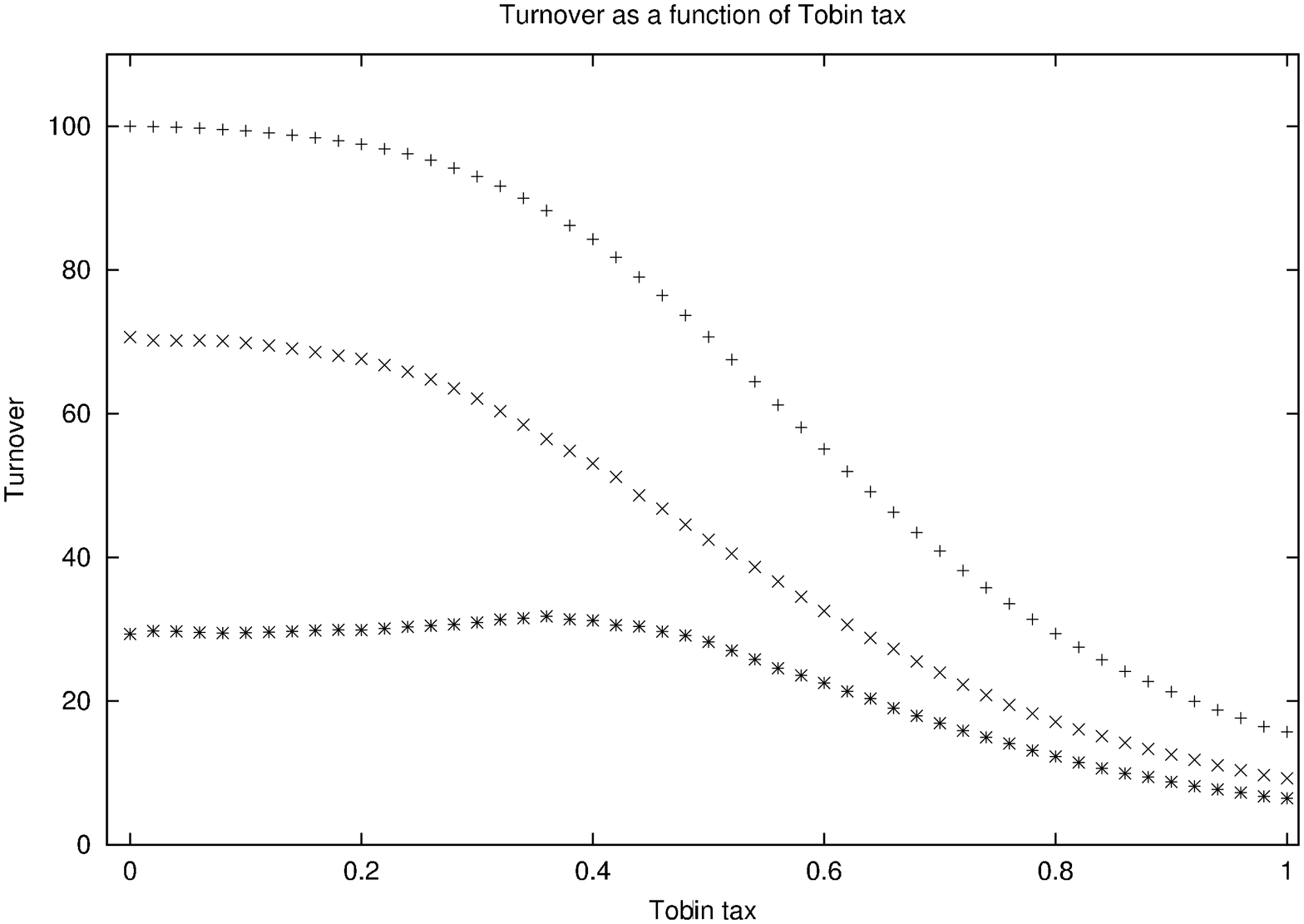}
\end{center}
\caption{
a) Tobin tax to get maximal profit as a function of Producers. b) Turnover as a 
function of Tobin tax with 1\% Producers.}
\end{figure}

\item Turnover as a function of Tobin tax\\[0.1cm]

In fig.13b we take 1\% Producers and maxwin= 5\%. We show the way
 the turnover will decrease if we introduce a Tobin Tax. The ``+'' sign is the 
turnover we get by buying and selling, ``x'' presents just the turnover by buying 
and ``*'' shows the turnover we get only by selling. To get the maximal profit 
we have to take a Tobin tax of 0.48\% under condition of 1\% Producers (we call 
this value $T_{max}$. $T_{max}$ is Ttp for one percent Producers). We see here 
that we have there still a turnover of 73\%.\\[0.1cm]

\item Return as a function of time\\[0.2cm]
We calculate the return as a function of time as described in the first 
modification of the Cont-Bouchaud model. We take 1\% Producers and 5\% maxwin.
The first line of fig.14 shows the behaviour for 0\% Tobin tax, the next line 
for 1\% Tobin tax and the line which fluctuate about zero (even for t equal 1) 
is for 0.48\% Tobin tax. We see here and with the following tabular that the 
prices will less fluctuate at the value of Tobin tax where government profit 
has a maximum.
\begin{figure}[!]
\begin{center}
\includegraphics[angle=-90,scale=0.5]{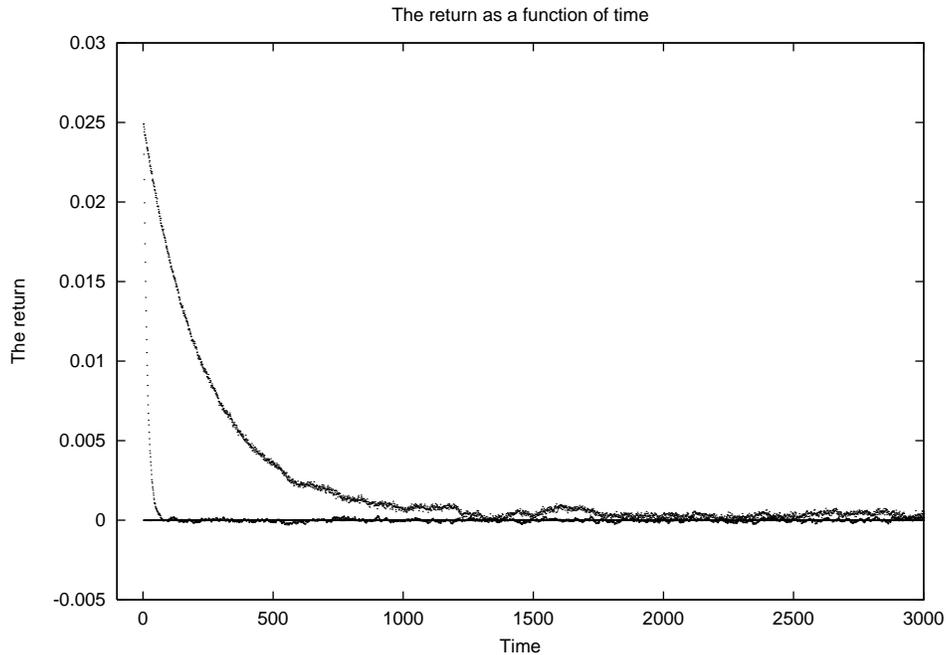}
\end{center}
\caption{
 The return as a function of time.}
\end{figure}
We got the following tabular by maxwin= 5\%. Tt means ``Tobin tax'', STFZ is 
the first time step where the return starts to fluctuate about zero, sd means 
``standard deviation'' and we find in the last column the value of $r_-$ at t=1 
which we call $r_-1$.

\begin{tabular}{|l|l|l|l|l|l|}
\hline
Tt & actual returns & STFZ & average & sd & $r_-1$ \\
\hline
0 & ($-$0.04\%)-0.12\% & 1000 &  $-$4.1E($-$2)&2.6E($-$2) & 2.49\%\\
\hline
0.01 & ($-$0.1\%)-0.08\% & 1000 & $-$1.7E($-$3) & 2.7E($-$2) & 2.49\%\\
\hline
0.16 & ($-$0.09\%)-0.09\%& 1000 & 1.4E($-$3) &  2.7E($-$2) & 2.49\%\\
\hline
0.2 & ($-$0.09\%)-0.11\% & 1000 & $-$7.9E($-$3)& 2.9E($-$2)&2.49\%\\
\hline
0.4 & ($-$0.09\%)-0.11\% & 1000 & $-$4.7E($-$3)&2.4E($-$2)&2.46\%\\
\hline
0.48 & ($-$0.007\%)-0.007\%& 0 &  $-$1.7E($-$4) &1.9E($-$3)& 0.001\%\\
\hline
0.5 &($-$0.007\%)-0.007\% & 0 & $-$1.7E($-$4)&1.9E($-$3)& 0.001\%\\
\hline
1.0 &($-$0.03\%)-0.03\% & 70 &  $-$2.7E($-$4)&9.2E($-$3)& 2.49\%\\
\hline
\end{tabular}

The returns as a function of time have a minimum of fluctuations for $T=T_{max}$. 
This shows that one get both, on the one hand maximal government profit and on the 
other hand a minimum of return fluctuations which is important to be able to reduce 
speculation.

The following tabular we get for maxwin= 50\%. Remember that in this case the 
maximum of government profit is above 1.5\% Tobin tax.

\begin{tabular}{|l|l|l|l|l|l|}
\hline
Tt & actual returns & STFZ & average & sd &  $r_-1$\\
\hline
0 & ($-$0.3\%)-1.2\%& 1000 &4.1E($-$1)&2.6E($-$1)&24.9\%\\
\hline
0.5 & ($-$0.8\%)-0.8\%& 1000 &$-$4.3E($-$2)&2.4E($-$1)&24.9\%\\
\hline
\end{tabular}

You see that here the return fluctuations are much higher than for maxwin =5\%.

\item The Return Histogram\\[0.2cm]

Fig.15a: shows from outside to inside the behaviour for 0\% Tobin tax and 0.5\% 
Tobin tax for an activity of a= 1\%. Fig.15b shows the price histogram for an 
activity of 50\% and a Tobin tax of 1.8\%.
Unfortunately we have not a realistic price histogram for an activity of 50\% 
and a Tobin tax below 1.8\%.

\begin{figure}[!]
\begin{center}
\includegraphics[angle=-90,scale=0.5]{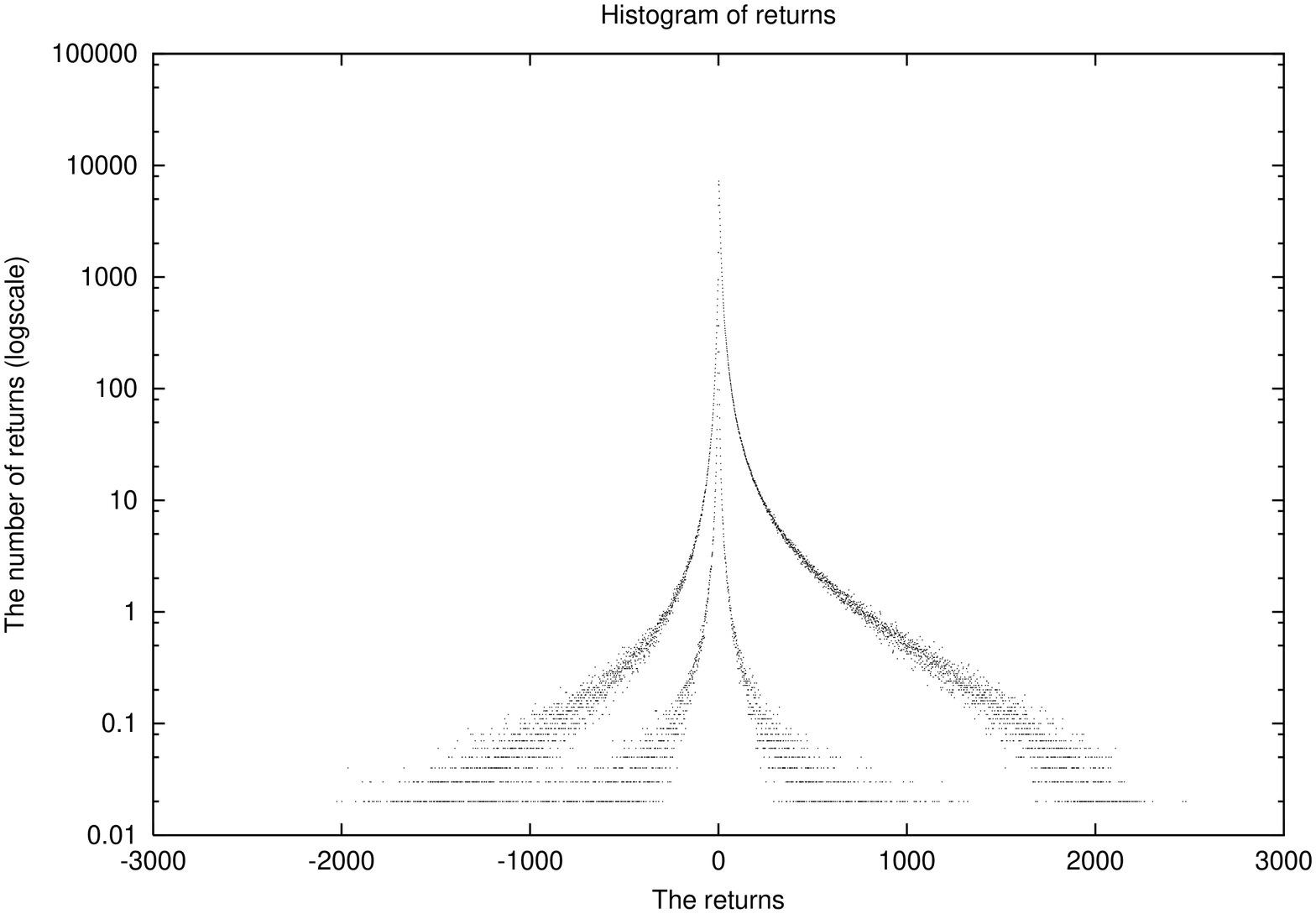}
\includegraphics[angle=-90,scale=0.5]{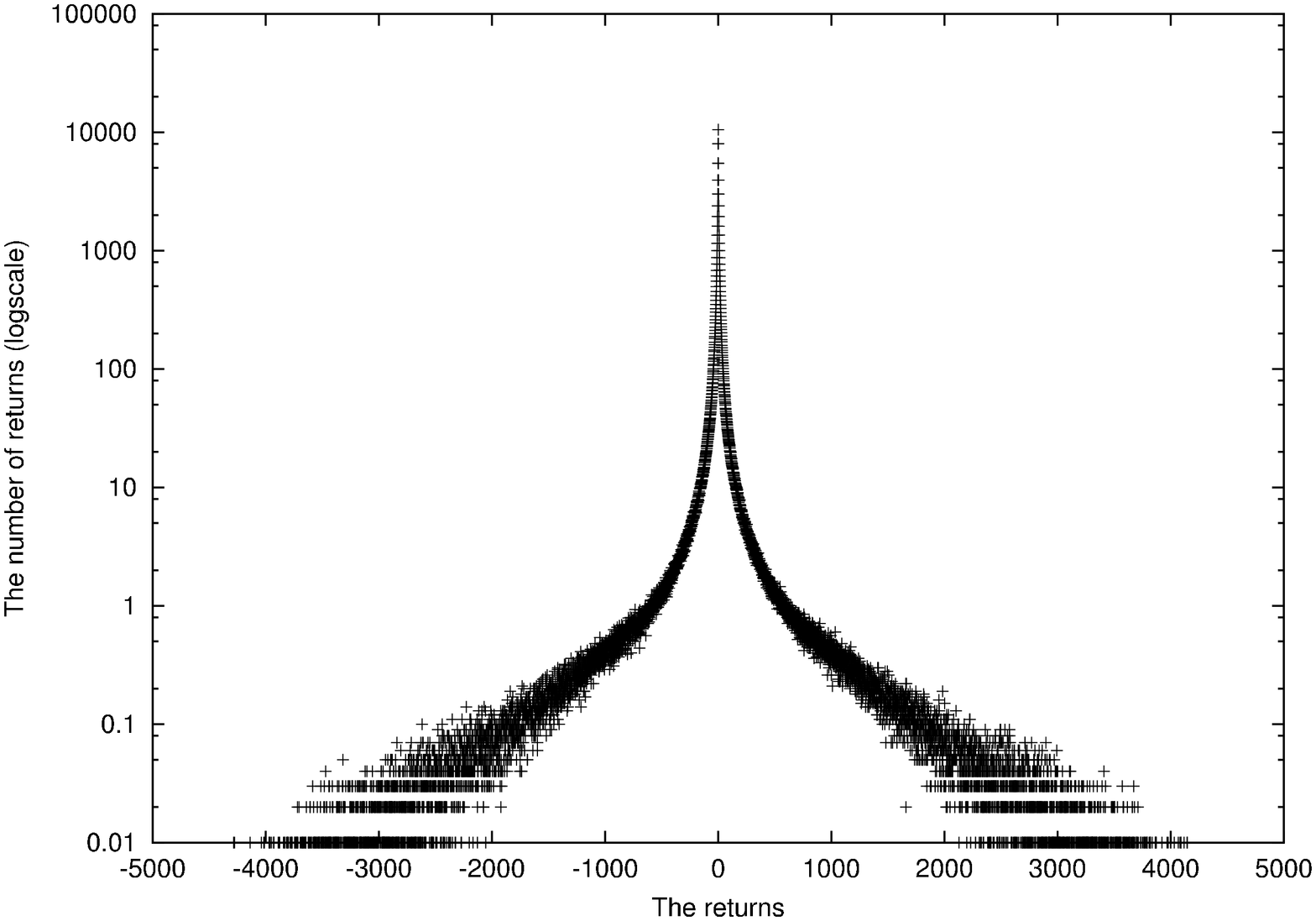}
\end{center}
\caption{a) Histogram of returns with a=1\%.}
\end{figure}

\begin{center}
\begin{tabular}{|l|l|l|}
\hline
\multicolumn{3}{|c|}{Kurtosis}\\
\hline
activity   & Tobin tax &Kurtosis \\
\hline
50\% & 1.8 & 160\\
\hline
50\% &2 &170\\
\hline 
50\% &10&109\\
\hline
1\% &0 & 103\\                 
\hline
1\%& 0.5 & 5130\\
\hline
1\%&1&5271\\
\hline
1\% & 10 & 5604\\
\hline
\end{tabular}
\end{center} 

\item Number of former speculating and still active speculating investors\\[0.2cm]
\begin{figure}[!]
\begin{center}
\includegraphics[angle=-90,scale=0.31]{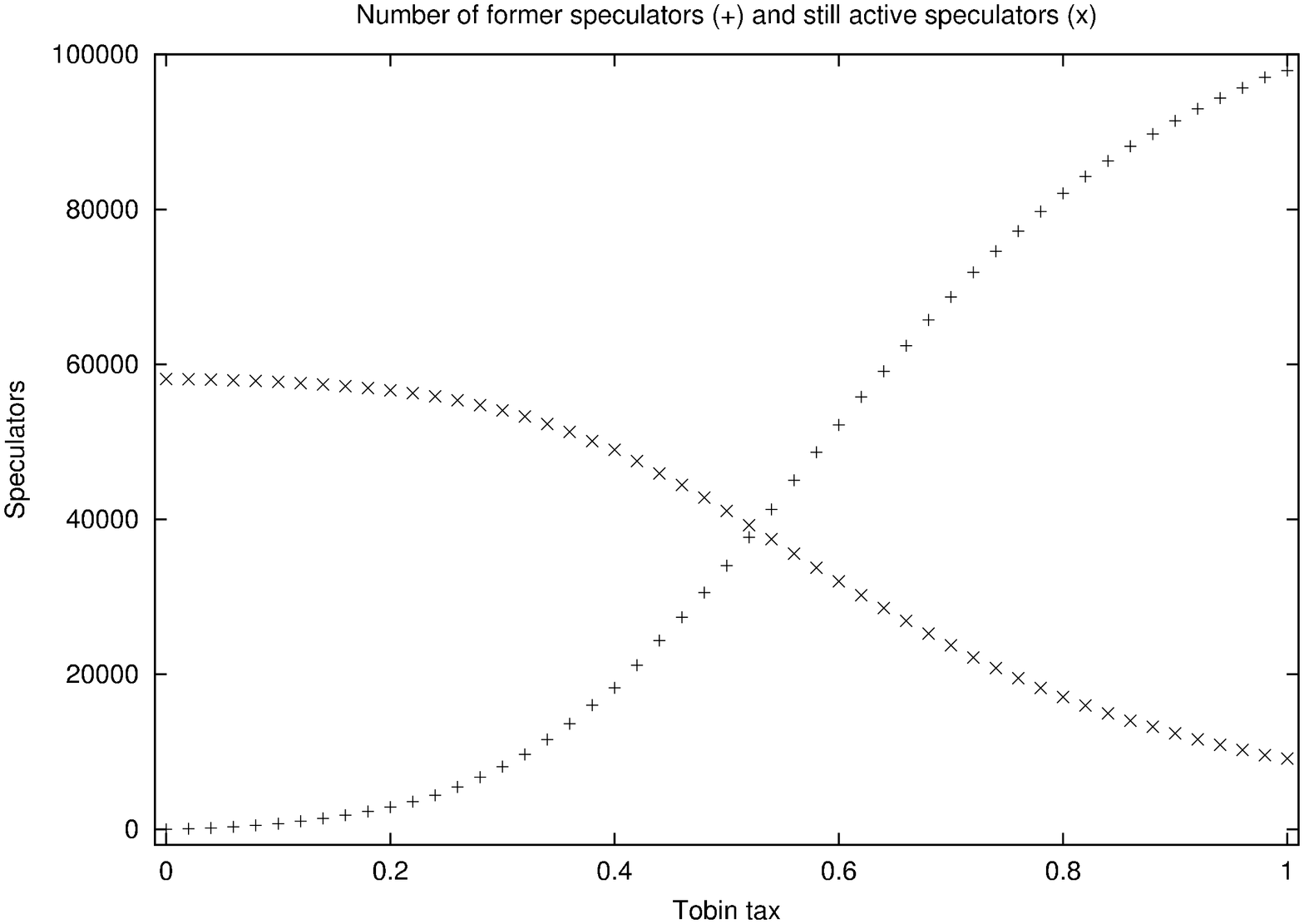}
\includegraphics[angle=-90,scale=0.31]{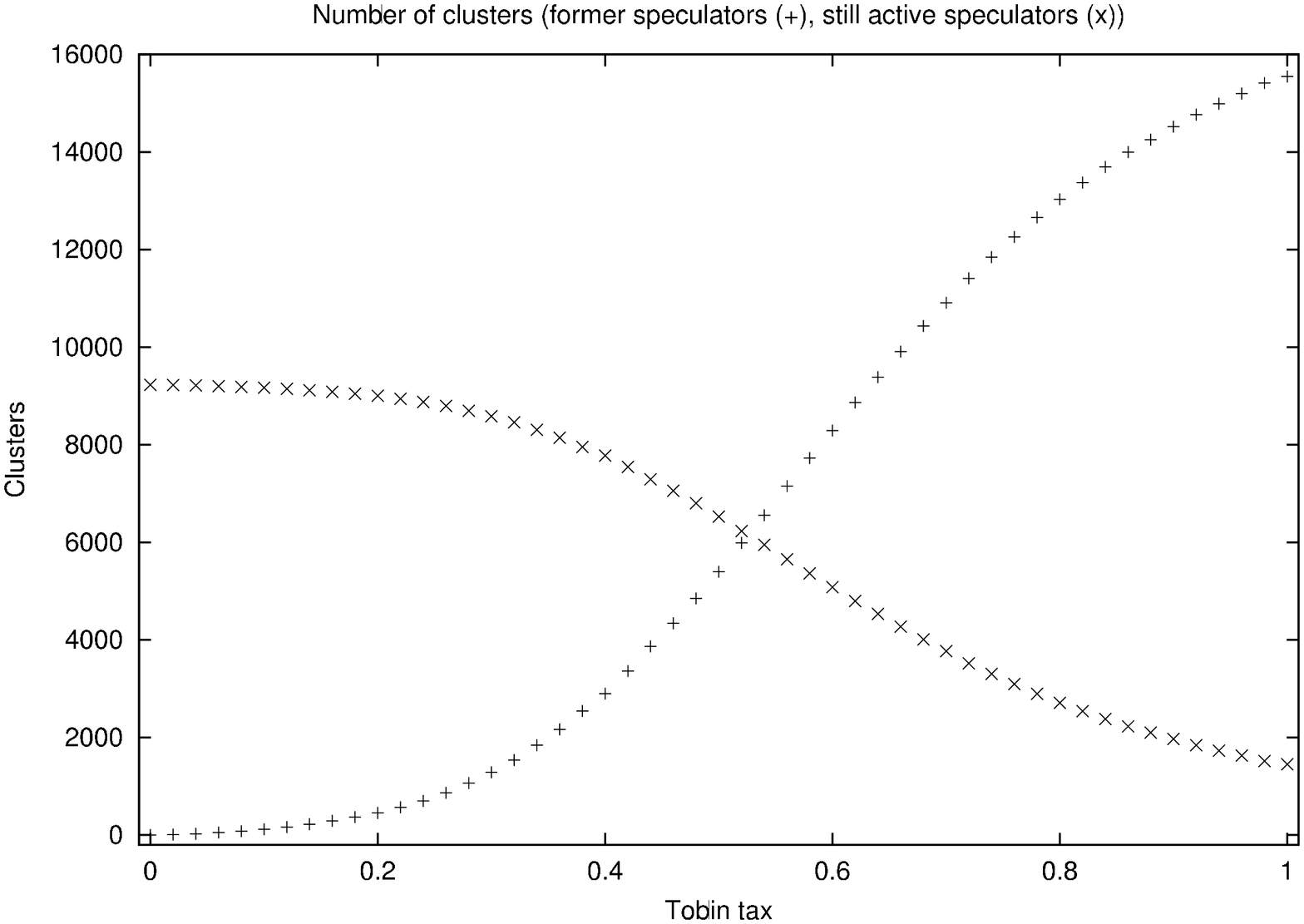}
\includegraphics[angle=-90,scale=0.31]{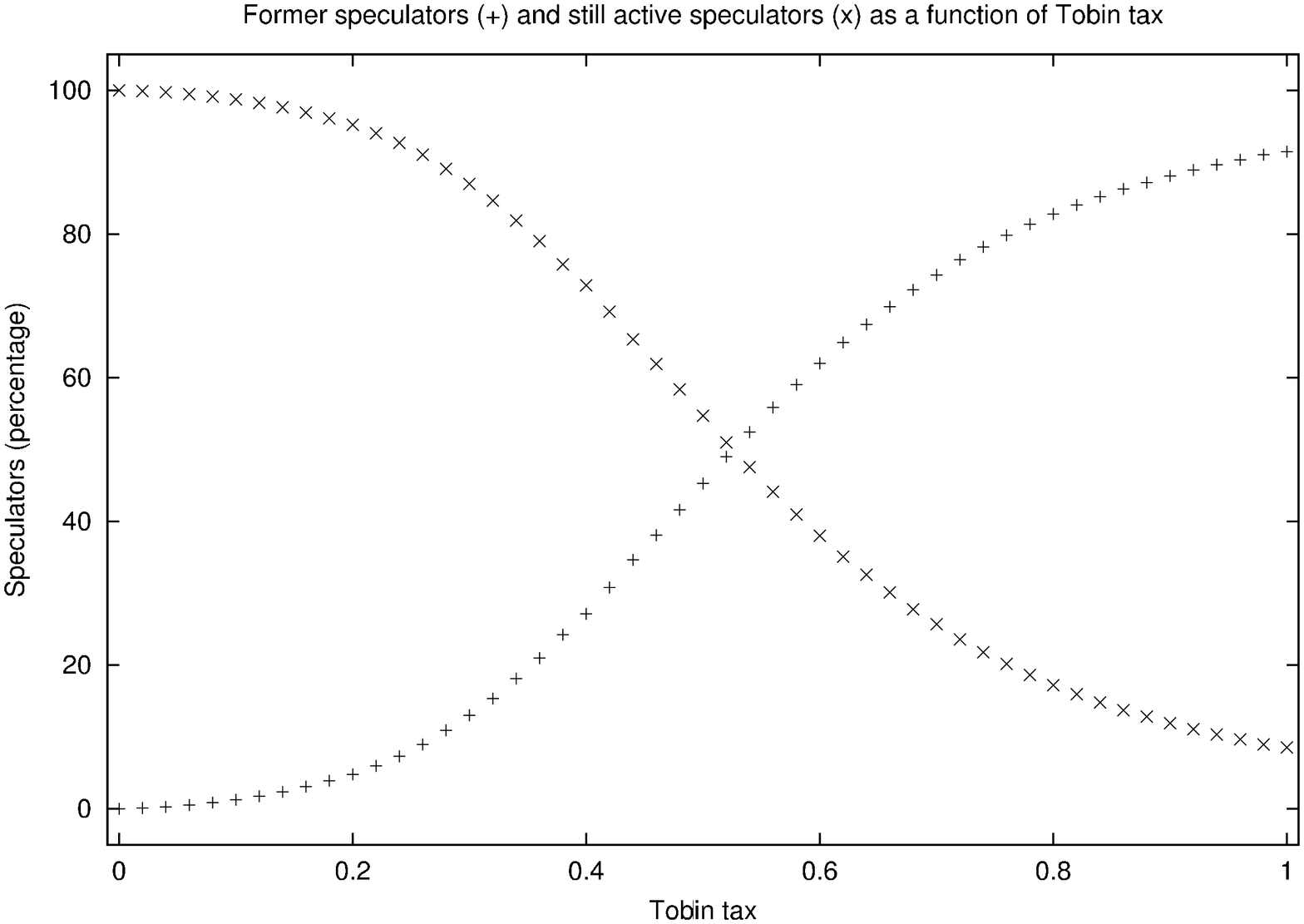}
\end{center}
\caption{ a) Number of former speculators (+) and still active speculators (x). 
b) Number of clusters (former speculators (+) and still active speculators (x)). 
c) Number of former speculators or still active speculators (percentage).}
\end{figure}

For Fig.16, we sum up the number of investors who are former speculators (+) or 
still active speculators (x) about 500 time steps for different values of Tobin 
tax rate. We define former speculators in the following way: The former speculators 
are those people who don't trade because they do not fulfill one of the conditions 
$r_-$ greater than Tobin tax or  $r_-$ less than ($-$ Tobin tax). Here is maxwin= 5\% 
and Producers= 1\%. The curves intersect at 0.53\% Tobin tax. At $T_{max}$ (0.48\% 
Tobin tax) we have here 42\% former speculators and 58\% still active speculators.\\
\end{itemize}
\begin{center}
 \textbf{Summary}
\end{center}
In conclusion, the first modification of the Cont-Bouchaud model towards an 
introduction of a Tobin tax does not lead to the desired maximum value for profit 
as a function of Tobin tax rate.  Instead, it discourages most speculators (96\%) 
if it becomes of the order of 0.5\%. \\
The differentiation between these two models (RM and SCBM) does not lead to 
different values one should propose for a Tobin tax rate. But when we multiply 
the returns of Cont-Bouchaud model by 10, the Tobin tax rate needs to be 
multiplied by 1.33.\\[0.2cm]
\begin{tabular}{|l|l|l|l|}
\hline
\multicolumn{4}{|c|}{Results about Tobin tax which one should propose}\\
\hline
maxwin &return &tax rate (RM) &tax rate (SCBM)\\
\hline
5\% & $\pm0.8\%$ &0.12\% & 0.12\%\\
\hline                       
50\% & $\pm8\%$ &0.16\% &0.16\%\\
\hline
\end{tabular}  \\[0.2cm]

\begin{tabular}{|l|l|l|}
\hline
\multicolumn{3}{|c|}{Kurtosis}\\
\hline
activity &Tobin tax or model  &Kurtosis      \\
\hline
50\% & original Cont-Bouchaud model &0.58\\
\hline 
10\% &original Cont-Bouchaud model & 4.96\\                 
\hline
1\%& original Cont-Bouchaud model &54.72\\
\hline
\hline
50\% &$r_-\ge0\%$ & 3.71\\
\hline
50\% &$r_-\ge0.01\%$ &8.38\\
\hline
50\% &$r_-\ge0.16\%$ & 56.76\\
\hline
\end{tabular}  \\[0.2cm]
The second modification leads to a desired maximum for government profit, 
sufficient turnover and damped return oscillations but not to a realistic 
return histogram for maximal activity and a Tobin tax below 1.8\%.
On the other hand it is a further success of this modification that when we 
work at $T_{max}$ we discourage 42\% of the speculators.
\newpage
\begin{center}
\textbf{Acknowledgements}
\end{center}
I would like to thank Prof. Stauffer for advice             .\\
\begin{center}
\textsl{\textbf{ References}}
\end{center}
\begin{enumerate}
\item Microsoft Encarta 2001 encyclopaedia: Bretton Woods\\
http://www.microsoft.com.england/
\item  Anja Osterhaus, Kai Mosebach, Peter Wahl and Peter Waldow:\\
Spekulieren regulieren: Kapital braucht Kontrolle, \\
WEED und Kairos Europa\\
http://www.weedbonn.org/
\item Danny Cassimon: Financing sustainable development 
using a feasible Tobin tax,\\
 Journal of International Relations and Development \textbf{4}, 157 (2001)\\ 
http://www.ub.uni-koeln.de/usb/dienste/ezb
\item Thomas Palley: Destabilizing speculation and the case for an International 
Currency Transaction tax,\\
 Challenge, \textbf{44}, 70 (2001)
http://www.ub.uni-koeln.de/usb/dienste/ezb
\item Peter Wahl and Peter Waldow: Devisenumsatzsteuer: Ein Konzept mit Zukunft, 
WEED, http://www.weedbonn.org/
\item Rama Cont and Jean-Philippe Bouchaud: Herd behavior and aggregate 
fluctuations in financial markets, Macroeconomic Dynamics \textbf{4}, 170 (2000)
\item Dietrich Stauffer: Monte-Carlo-Simulation mikroskopischer Boersenmodelle, 
Physikalische Bl\"atter,  \textbf{55}, 49 (1999)
\item Dietrich Stauffer: Percolation models of financial market dynamics,\\
 Adv.Compl.Syst. 4,19(2001) \\
\end{enumerate}

\begin{center}
\textsl{\textbf{ For Further Reading}}
\end{center}
\begin{itemize}
\item Howard M. Wachtel: Tobin and other global taxes,\\ 
Review of International Political Economy \textbf{7}, 335 (2000)\\
http://www.ub.uni-koeln.de/usb/dienste/ezb
\item Robert W. Dimand and Mohammed H.I. Dore\\
Keynes   casino capitalism, Bagehot's international currency,\\
and the Tobin tax: historical notes on preventing currency fires\\
Journal of Post Keynesian Economics \textbf{22}, 515, (2000)

\end{itemize}
\end{document}